\documentclass[12pt,a4paper]{article}

\usepackage[bulletsep]{collref}
\usepackage{amsmath,amssymb,graphicx}
\usepackage{pst-all}
\usepackage{bbm}
\makeatletter \@addtoreset{equation}{section} \makeatother

\makeatletter
\let\old@startsection=\@startsection
\let\oldl@section=\l@section
\renewcommand{\@startsection}[6]{\old@startsection{#1}{#2}{#3}{#4}{#5}{#6\mathversion{bold}}}
\renewcommand{\l@section}[2]{\oldl@section{\mathversion{bold}#1}{#2}}
\makeatother

\makeatletter
\let\old@makecaption=\@makecaption
\def\@makecaption{\small\old@makecaption}
\makeatother

\renewcommand{\leq}{\leqslant}
\renewcommand{\geq}{\geqslant}

\begin{document}

\thispagestyle{empty}
\begin{flushright}\footnotesize
\texttt{ITEP-TH-12/10}\\
\texttt{LPTENS-10/12}\\
\texttt{UUITP-05/10} \vspace{0.8cm}
\end{flushright}

\renewcommand{\thefootnote}{\fnsymbol{footnote}}
\setcounter{footnote}{0}

\begin{center}
{\Large\textbf{\mathversion{bold} Strings on Semisymmetric
Superspaces}\par}

\vspace{1.5cm}

\textrm{K.~Zarembo\footnote{Also at ITEP, Moscow, Russia}}
\vspace{8mm}

\textit{CNRS -- Laboratoire de Physique Th\'eorique,
\'Ecole Normale Sup\'erieure\\
24 rue Lhomond, 75231 Paris, France }\\
\texttt{Konstantin.Zarembo@lpt.ens.fr} \\ \vspace{3mm} and
\vspace{3mm}

\textit{Department of Physics and Astronomy, Uppsala University\\
SE-751 08 Uppsala, Sweden}\\
\vspace{3mm}

%%%%%%%%

\par\vspace{1cm}

\textbf{Abstract} \vspace{5mm}

\begin{minipage}{14cm}
Several string backgrounds which  arise in the AdS/CFT
correspondence are described by integrable sigma-models. Their
target space is always a $\mathbbm{Z}_4$ supercoset (a
semi-symmetric superspace). Here we list all semi-symmetric cosets
which have zero beta function and central charge $c\leq 26$ at one
loop in perturbation theory.
\end{minipage}

\end{center}

\vspace{0.5cm}

%%%%%%%%%%%%%%%%%%%%%%%%%%%%%%%%%%%%%%%%%%%%%%%%%%%%%%%%%%%%%%%%%%%%%%%%%%%%%%%%

\newpage
\setcounter{page}{1}
\renewcommand{\thefootnote}{\arabic{footnote}}
\setcounter{footnote}{0}

%%%%%%%%%%%%%%%%%%%%%%%%%%%%%%%%%%%%%%%%%%%%%%%%%%%%%%%%%%%%%%%%%%%%%%%%%%%%%%%%
\section{Introduction}

A semi-symmetric superspace is a coset of a supergroup which
possesses an additional $\mathbbm{Z}_4$ symmetry \cite{Serganova},
thus generalizing the notion of ordinary, $\mathbbm{Z}_2$ invariant
symmetric space. Sigma-models on semi-symmetric superspaces possess
a number of interesting properties. Perhaps the main motivation to
study them comes from the AdS/CFT duality. The holographic duals of
superconformal field theories in diverse dimensions are string
theories on Anti-de-Sitter backgrounds with Ramond-Ramond fluxes. In
many cases (and certainly in all maximally symmetric cases), the
worldsheet sigma-models on such backgrounds are $\mathbbm{Z}_4$
cosets \cite{Berkovits:1999zq}. The best known example is the
Green-Schwarz string action on $AdS_5\times S^5$
\cite{Metsaev:1998it}, which is a $\mathbbm{Z}_4$ coset of
$PSU(2,2|4)$, the superconformal group in four
dimensions\footnote{The manifestly $\mathbbm{Z}_4$-invariant form of
the Metsaev-Tseytlin action is given in \cite{Roiban:2000yy}.}. One
can define a Green-Schwarz-type sigma-model on any $\mathbbm{Z}_4$
coset. The $\mathbbm{Z}_4$ symmetry plays a crucial role in this
construction by yielding the fermionic Wess-Zumino term in the
sigma-model action \cite{Berkovits:1999zq}.

A remarkable property of the Green-Schwarz-type $\mathbbm{Z}_4$
cosets is their classical integrability \cite{Bena:2003wd}, which
parallels integrability of bosonic sym\-met\-ric-space sigma-models
\cite{Eichenherr:1979ci}. A Lax representation of the equations of
motion in semi-symmetric cosets can be constructed using the
$\mathbbm{Z}_4$ symmetry in a uniform, purely algebraic
way\footnote{The original construction of \cite{Bena:2003wd} for
$PSU(2,2|4)/SO(4,1)\times SO(5)$ relies only on the $\mathbbm{Z}_4$
decomposition of the symmetry algebra and thus applies to any
semi-symmetric coset.}. Perhaps this is why integrability arises in
the AdS/CFT correspondence.

All semi-symmetric superspaces are classified \cite{Serganova}, and
one can scan the list of the $\mathbbm{Z}_4$ cosets for potentially
interesting integrable models, in  particular for integrable string
backgrounds. To be a string background, a $\mathbbm{Z}_4$ coset must
satisfy two additional conditions: its beta function should vanish
and it should have central charge $c=26$.

After reviewing the construction of the Green-Schwarz-type
sigma-model on a semi-symmetric superspace, we will compute its beta
function and central charge at one loop following
\cite{Polyakov:1975rr,Polyakov:2004br,Adam:2007ws}. Then we will
list all cosets that satisfy the beta-function and the central
charge constraints.

%%%%%%%%%%%%%%%%%%%%%%%%%%%%%%%%%%%%%%%%%%%%%%%%%%%%%%%%%%%%%%%%%%%%%%%%%%%%%%%%
\section{Sigma Model}

A coset $G/H_0$ of a supergroup $G$ is a semi-symmetric superspace
if it is invariant under a $\mathbbm{Z}_4$ symmetry, generated by a
linear automorphism $\Omega $ of the Lie algebra of $G$, $\Omega
:\mathfrak{g}\rightarrow \mathfrak{g},~\Omega ([X,Y])=[\Omega
(X),\Omega (Y)],~\Omega ^4={\rm id}$. The superalgebra
$\mathfrak{g}$ then admits a $\mathbbm{Z}_4$ decomposition:
\begin{equation}\label{}
 \mathfrak{g}=\mathfrak{h}_0\oplus\mathfrak{h}_1\oplus\mathfrak{h}_2\oplus\mathfrak{h}_3,
\end{equation}
which is consistent with the (anti-)commutation relations:
$[\mathfrak{h}_n,\mathfrak{h}_m\}\subset
\mathfrak{h}_{(n+m)\!\!\mod\! 4}$. The subspace $\mathfrak{h}_n$
consists of the elements of $\mathfrak{g}$ with the $\mathbbm{Z}_4$
charge $n$:
\begin{equation}\label{}
 \Omega (\mathfrak{h}_n)=i^n\mathfrak{h}_n.
\end{equation}
The denominator subalgebra of a semi-symmetric coset is the
$\mathbbm{Z}_4$-invariant subspace $\mathfrak{h}_0$. The fermion
number $F$ is the $\mathbbm{Z}_4$ charge $\!\!\mod 2$: the bosonic
subalgebra of $\mathfrak{g}$ is $\mathfrak{h}_0\oplus\mathfrak{h}_2$
and all of the odd generators belong to either $\mathfrak{h}_1$ or
$\mathfrak{h}_3$.

The worldsheet embedding in $G/H_0$ is parameterized by a coset
representative $g(x)\in G$, subject to gauge transformations
$g(x)\rightarrow g(x)h(x)$ with $h(x)\in H_0$. The global $G$-valued
transformations act on $g(x)$ from the left: $g(x)\rightarrow
g'g(x)$. The  action of the sigma-model can be written in terms of
the $\mathbbm{Z}_4$ decomposition of the left-invariant current
$g^{-1}\partial _\mu g$:
\begin{equation}\label{cur}
 J_\mu =g^{-1}\partial _\mu g=J_{\mu\,0 }+J_{\mu\,1 }+J_{\mu\,2
 }+J_{\mu\,3 }.
\end{equation}
The $\mathfrak{h}_0$ component of the current transforms as a
connection under gauge transformations: $J_{\,\mu }\rightarrow
h^{-1}J_{\mu\,0 }h+h^{-1}\partial _\mu h$. The other three
components transform as matter fields in the adjoint: $J_{\mu
\,1,2,3}\rightarrow h^{-1}J_{\mu \,1,2,3}h$.

The action must be gauge invariant and $\mathbbm{Z}_4$-symmetric. By
power counting, the only possible terms\footnote{These terms
describe coupling to the metric and to the RR fields. In certain
cases it should be possible to switch on the B-field (the theta-term
in the sigma-model action) or its field strength (the bosonic
Wess-Zumino term). For example, if the denominator of the coset
contains a $U(1)$ factor, it is possible to add a theta-term
$i\vartheta\varepsilon ^{\mu \nu }\partial _\mu J^{U(1)}_{\nu
\,0}$.} are $J_2J_2$ and $J_1J_3$. A particularly interesting case,
and the one that we will consider here is\footnote{We consider the
Euclidean worldsheet, which is why the second term in the Lagrangian
is multiplied by $i$. After the Wick rotation the action becomes
real.}
\begin{equation}\label{action}
 S=\frac{1}{2\kappa^2 }\int_{}^{}d^2x\,\mathop{\mathrm{Str}}
 \left(\sqrt{h}h^{\mu \nu }J_{\mu\,2 }J_{\nu \,2}+i\varepsilon ^{\mu \nu }
 J_{\mu\,1 }J_{\nu\,3 }\right).
\end{equation}
The "supertrace" $\mathop{\mathrm{Str}}(\cdot \,\cdot )$ denotes the
$G$ and $\mathbbm{Z}_4$ invariant bilinear form on $\mathfrak{g}$,
and $\kappa $ is the sigma-model coupling ($\kappa^2 =2\pi \alpha
'/R^2$, where $R$ is the radius of $G/H_0$). The equations of motion
for this action admit a Lax representation \cite{Bena:2003wd} making
the world-sheet sigma-model classically integrable.

The expansion of the Lagrangian in (\ref{action}) around $g=1$ (the
flat-space limit) has the form $\partial X\partial X+\bar{\theta
}\partial X\partial \theta $ typical for the Green-Schwarz
superstring \cite{Green:1983wt}. And indeed the Green-Schwarz action
on many AdS backgrounds can be described as (\ref{action}) for
various $\mathbbm{Z}_4$ cosets
\cite{Metsaev:1998it,Rahmfeld:1998zn,Park:1998un,Zhou:1999sm,Metsaev:2000mv,Verlinde:2004gt,Polyakov:2004br,Chen:2005uj,Adam:2007ws,Hatsuda:2008xa,Arutyunov:2008if,Stefanski:2008ik,Babichenko:2009dk}.
Just like the ordinary Green-Schwarz action, (\ref{action}) may
possess local fermionic kappa-symmetries which, in effect, means
that some of the fermion dimensions are unphysical and have to be
removed by an appropriate gauge fixing prior to quantization. The
rank of the kappa-symmetry depends on the structure of the coset and
will be computed in sec.~\ref{conforms} for all sigma-models with
the vanishing one-loop beta-function.

To illustrate these points and to set up the stage for the
subsequent one-loop calculations, let us expand the action
(\ref{action}) around an arbitrary bosonic background\footnote{The
background-field calculations have been done for the
Green-Schwarz-type cosets \cite{Polyakov:2004br}, as well as for
many related pure-spinor type sigma-models
\cite{Berkovits:1999zq,Vallilo:2002mh,Kagan:2005wt,Puletti:2006vb,Adam:2007ws,Mazzucato:2009fv}.}
$\bar{g}(x)$, introducing the following notations for the background
currents:
\begin{eqnarray}\label{}
 \left(\bar{g}^{-1}\partial _\mu \bar{g}\right)_0&=&A_\mu ,\nonumber \\
 \left(\bar{g}^{-1}\partial _\mu \bar{g}\right)_2&=&K_\mu.
\end{eqnarray}
Here $A_\mu $ is the background gauge field. We will denote by
$D_\mu $ the background covariant derivative: $D_\mu =\partial _\mu
+[A_\mu ,\cdot ]$, and by $F_{\mu \nu }$ the background field
strength: $F_{\mu \nu }=\partial _\mu A_{\nu }-\partial _\nu A_\mu
+[A_\mu ,A_\nu ]$. The currents $A_\mu $ and $K_\mu $ are assumed to
satisfy the classical equations of motion:
\begin{eqnarray}\label{eqmo}
 && [K_\mu ,K_\nu ]+F_{\mu \nu }=0, \nonumber \\
 && D_\mu K_\nu -D_\nu K_\mu =0, \nonumber \\
 && \nabla_\mu K^\mu =0,
\end{eqnarray}
where $\nabla_\mu K^\nu  =D_\mu K^\nu  +\Gamma^\nu _{\mu \lambda
}K^\lambda  $ and $\Gamma ^\nu _{\mu \lambda }$ are the Christoffel
symbols of the worldsheet metric. The first two equations are
identities that follow from the flatness of the current
$\bar{g}^{-1}\partial _\mu \bar{g}$. The equations of motion for the
metric are the Virasoro constraints:
\begin{equation}\label{Viras1}
 h^{\mu \nu }\mathop{\mathrm{Str}}K_{\pm\,\mu} K_{\pm\,\nu} =0,
\end{equation}
where $K_{\pm\,\mu }$ are the chiral components of $K_\mu $:
\begin{eqnarray}\label{chirvec}
 K_{\pm\, \mu }=\frac{1}{2}\left(\delta _\mu ^\nu \pm\frac{i}{\sqrt{h}}\,
 h_{\mu \lambda }\varepsilon ^{\lambda  \nu }
 \right)K_\nu .
\end{eqnarray}

In order to expand around the classical background $\bar{g}(x)$ we
choose the coset representative in the form
\begin{equation}\label{core}
 g=\bar{g}\,{\rm e}\,^{\kappa X},
\end{equation}
where $X\in\mathfrak{h}_1\oplus\mathfrak{h}_2\oplus\mathfrak{h}_3$.
Under gauge transformation that also act on the background field:
$\bar{g}\rightarrow \bar{g}h$, $X$ transforms in the adjoin:
$X\rightarrow h^{-1}Xh$. It is straightforward to plug the coset
representative (\ref{core}) into the action and expand the latter in
the powers of the coupling $\kappa $. The current (\ref{cur})
expands as
\begin{equation}\label{curexp}
 J_\mu =A_\mu +K_\mu +\frac{1-\,{\rm e}\,^{-\kappa \mathop{\mathrm{ad}}X}}
 {\mathop{\mathrm{ad}}X}\,\mathcal{D}_\mu X
 =
 A_\mu +K_\mu +\kappa \mathcal{D}_\mu X-\frac{\kappa ^2}{2}[X,\mathcal{D}_\mu X]
 +\ldots ,
\end{equation}
where the long derivative $\mathcal{D}_\mu $ is defined by
\begin{equation}\label{}
 \mathcal{D}_\mu =\partial _\mu +[\bar{g}^{-1}\partial _\mu \bar{g},\cdot ]
 =D_\mu +[K_\mu ,\cdot ].
\end{equation}
Unlike the covariant derivative $D_\mu $, which commutes with the
$\mathbbm{Z}_4$ grading, the long derivative $\mathcal{D}_\mu $ does
not have definite $\mathbbm{Z}_4$ charge. Thus, $(D_\mu X)_n=D_\mu
X_n$ for any $n$ and $(\mathcal{D}_\mu X)_2=D_\mu X_2$, but
$(\mathcal{D}_\mu X)_{1,3}=D_\mu X_{1,3}+[K_\mu ,X_{3,1}]$.

Plugging the expansion (\ref{curexp}) into the action (\ref{action})
and using the identities
\begin{eqnarray}\label{}
 && \varepsilon ^{\mu \nu }D_\mu D_\nu =-\varepsilon ^{\mu \nu
 }\mathop{\mathrm{ad}}K_\mu \mathop{\mathrm{ad}}K_\nu, \nonumber \\
 && \varepsilon ^{\mu \nu }[D_\mu,\mathop{\mathrm{ad}}K_\nu  ]=0,\nonumber \\
 && [D_\mu ,\sqrt{h}h^{\mu \nu }\mathop{\mathrm{ad}}K_\nu ]=0,\nonumber
\end{eqnarray}
which follow from the equations of motion (\ref{eqmo}), one can
bring the quadratic part of the action to the form
\begin{eqnarray}\label{action2}
 S&=&\bar{S}+\int_{}^{}d^2x\,\sqrt{h}h^{\mu \nu }\mathop{\mathrm{Str}}
 \left(
 \frac{1}{2}\,D_\mu X_2D_\nu X_2-\frac{1}{2}\,[K_\mu ,X_2][K_\nu ,X_2]
  \right. \nonumber \\ &&\left.
  +X_1\nabla_{+\,\mu }[K_{-\,\nu },X_1]
\vphantom{\frac{1}{2}}
 +X_3\nabla_{-\,\mu }[K_{+\,\nu },X_3]-2[K_{+\,\mu },X_3][K_{-\,\nu },X_1]
 \right)
 \nonumber \\
 &&+O(\kappa X^3),
\end{eqnarray}
where the chiral projections of a vector are defined in
(\ref{chirvec}). In the conformal gauge ($h_{\mu \nu }=\,{\rm
e}\,^{\phi }\delta _{\mu \nu }$),  the quadratic part of the
Lagrangian becomes:
\begin{eqnarray}\label{action2conf}
 \mathcal{L}_2&=&
 =\frac{1}{2}
 \mathop{\mathrm{Str}}\left(
 \bar{D}X_2DX_2-[\bar{K},X_2][K,X_2]\right. \nonumber \\ &&\left.
 +X_1D[\bar{K},X_1]+X_3\bar{D}[K,X_3]-2[K,X_3][\bar{K},X_1]
 \right),
\end{eqnarray}
where holomorphic and anti-holomorphic vector components are defined
as $D =D_1+iD_2$, $\bar{D}=D_1-iD_2$, and similarly for $K$.

The fermion fluctuations of the worldsheet couple to the background
currents, and if the currents vanish the fermion kinetic terms
vanish too. Even if the background currents do not vanish, the Dirac
operator may have zero modes, because the Lagrangian depends on
$X_1$ ($X_3$) only in the combination $[\bar{K},X_1]$ ($[K,X_3]$).
If $\bar{K}$ ($K$) has a non-empty commutant in $\mathfrak{h}_1$
($\mathfrak{h}_3$), the Lagrangian degenerates and simply does not
depend on the fermionic fluctuations in the corresponding
directions. This is a manifestation of the $\kappa $-symmetry, a
local fermion gauge invariance that has to be fixed in order to have
well-defined perturbation theory.

The most simple and natural way to fix the kappa-gauge is to set to
zero those components of $X_1$ and $X_3$ that drop out from the
action anyway. These components are proportional to the Lie algebra
generators from $\mathfrak{h}_1$ and $\mathfrak{h}_3$ which are
annihilated by the adjoint action of $\bar{K}$ or $K$. The rank of
the $\kappa $-symmetry is the number of such generators:
\begin{equation}\label{rankkappa}
 N_\kappa =\left.\dim\ker\mathop{\mathrm{ad}}K\right|_{\mathfrak{h}_3},\qquad
 N_{\tilde{\kappa }}
 =\left.\dim\ker\mathop{\mathrm{ad}}\bar{K}\right|_{\mathfrak{h}_1},
\end{equation}
where $K$ and $\bar{K}$ are sufficiently generic null elements of
$\mathfrak{h}_2$. The null condition follows from the Virasoro
constraints
\begin{equation}\label{null}
 \mathop{\mathrm{Str}}K^2=0=\mathop{\mathrm{Str}}\bar{K}^2.
\end{equation}

The number of zero modes $N_\kappa $ or $N_{\tilde{\kappa }}$ does
not depend on the particular choice of $K$ and $\bar{K}$ provided
that they are sufficiently generic. For special (non-generic)
classical solutions, the kappa-symmetry gauge condition may further
degenerate. This is known to happen in $AdS_4\times CP^3$
\cite{Arutyunov:2008if}. However these degenerate cases occur on the
surface of non-vanishing co-dimension in phase space. In the bulk of
the phase space (for generic classical solutions) the rank of the
kappa-symmetry is background independent, and is determined by the
structure constants of the Lie superalgebra $\mathfrak{g}$.

%%%%%%%%%%%%%%%%%%%%%%%%%%%%%%%%%%%%%%%%%%%%%%%%%%%%%%%%%%%%%%%%%%%%%%%%%%%%%%%%
\section{Beta Function and Central Charge}

To compute the central charge and the beta-function of the
sigma-model, we integrate out $X_n$, $n=1,2,3$ in (\ref{action2})
and study the dependence of the effective action on the background
currents and the 2d metric. The beta-function is determined by the
log-divergent contribution to the unique dimension two operator:
$\sqrt{h}h^{\mu \nu }\mathop{\mathrm{Str}}K_\mu K_\nu $. The central
charge is determined by the standard conformal anomaly. Since the
beta function and the central charge are governed by different terms
in the effective action, they can be computed separately. The beta
function arises from the insertions of the mass operators $K^2X_2^2$
and $K^2X_1X_3$ in the one-loop diagram and can be calculated in the
conformal gauge. The central charge arises due to the short-distance
anomaly in the fluctuation determinants and is insensitive to the
masses. In computing the central charge the masses can thus be
omitted, after which the Lagrangian (\ref{action2}) reduces to that
of the Green-Schwarz string in flat space in the semi-light-cone
gauge, the central charge for which was computed in
\cite{Carlip:1986cz,Kallosh:1988wv,Wiegmann:1989md}.

The one-loop effective action in the conformal gauge is
\begin{equation}\label{seff}
 S_{\rm eff}=\frac{1}{2}\,\mathop{\mathrm{Sp_2}}\ln\left(
 -\bar{D}D+\mathop{\mathrm{ad}}K\mathop{\mathrm{ad}}\bar{K}\right)
 -\frac{1}{2}\,\mathop{\mathrm{Sp'_{1\oplus 3}}}
 \begin{pmatrix}
    \mathop{\mathrm{ad}}K\mathop{\mathrm{ad}}\bar{K} & \bar{D}\mathop{\mathrm{ad}}K  \\
    D\mathop{\mathrm{ad}}\bar{K} & \mathop{\mathrm{ad}}\bar{K}\mathop{\mathrm{ad}}K  \\
 \end{pmatrix}.
\end{equation}
Here we used that $-D^\mu D_\mu +\mathop{\mathrm{ad}}K^\mu
\mathop{\mathrm{ad}}K_\mu
=-\bar{D}D+\mathop{\mathrm{ad}}K\mathop{\mathrm{ad}}\bar{K}$ because
of the identity $[D_\mu ,D_\nu ]=\mathop{\mathrm{ad}}F_{\mu \nu
}=-[\mathop{\mathrm{ad}}K_\mu ,\mathop{\mathrm{ad}}K_\nu ]$
satisfied by the background currents in virtue of the equations of
motion (\ref{eqmo}). The prime in $\mathop{\mathrm{Sp'_{1\oplus
3}}}$ means that the zero eigenvectors of
$\mathop{\mathrm{ad}}\bar{K}$ ($\mathop{\mathrm{ad}}K$) in
$\mathfrak{h}_1$ ($\mathfrak{h}_3$) should be omitted. They are
eliminated by fixing the kappa-symmetry gauge.

\begin{figure}[t]
\centerline{\includegraphics[width=5cm]{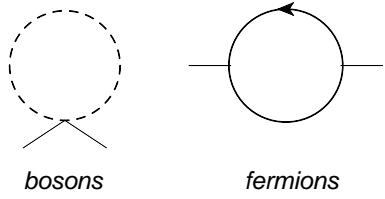}}
\caption{\label{dgr}\small The one-loop contribution to the
beta-function.}
\end{figure}

The log-divergent contribution to the beta-function comes from the
two diagrams in fig.~\ref{dgr}. The bosonic contribution is easy to
compute:
\begin{equation}\label{}
 \frac{1}{2}\int_{}^{}\frac{d^2p}{\left(2\pi \right)^2}\,\,\frac{1}{p^2}
 \int_{}^{}d^2x\,\,{\rm tr}_2\left(\mathop{\mathrm{ad}}K_\mu
 \right)^2=\frac{1}{4\pi }\,\ln\Lambda
 \int_{}^{}d^2x\,\,{\rm tr}_2\left(\mathop{\mathrm{ad}}K_\mu
 \right)^2.
\end{equation}

The fermion contribution requires more care because of the
kappa-symmetry projection. The Dirac operator in (\ref{seff}) can be
factorized as
\begin{eqnarray}\label{}
 \begin{pmatrix}
    \mathop{\mathrm{ad}}K\mathop{\mathrm{ad}}\bar{K} & \bar{D}\mathop{\mathrm{ad}}K  \\
    D\mathop{\mathrm{ad}}\bar{K} & \mathop{\mathrm{ad}}\bar{K}\mathop{\mathrm{ad}}K  \\
 \end{pmatrix}
 &=&
 \begin{pmatrix}
   \bar{D}  & \mathop{\mathrm{ad}}K  \\
     \mathop{\mathrm{ad}} \bar{K} & D  \\
 \end{pmatrix}
 \begin{pmatrix}
   0  & \mathop{\mathrm{ad}}K  \\
   \mathop{\mathrm{ad}} \bar{K} & 0  \\
 \end{pmatrix}
\nonumber \\
 &=&
\begin{pmatrix}
   0  & \mathop{\mathrm{ad}}K  \\
   \mathop{\mathrm{ad}} \bar{K} & 0  \\
 \end{pmatrix}
 \begin{pmatrix}
   D  & \mathop{\mathrm{ad}}K  \\
     \mathop{\mathrm{ad}} \bar{K} & \bar{D}  \\
 \end{pmatrix},\nonumber
\end{eqnarray}
where we used that
$[D,\mathop{\mathrm{ad}}\bar{K}]=0=[\bar{D},\mathop{\mathrm{ad}}K]$
due to the equations of motion. The Dirac operator acts on
$\mathfrak{h}_1\oplus\mathfrak{h}_3$, so the factor
$$
 \begin{pmatrix}
   0  & \mathop{\mathrm{ad}}K  \\
   \mathop{\mathrm{ad}} \bar{K} & 0  \\
 \end{pmatrix}
$$
is just the kappa-symmetry projector, up to proportionality factor.
The fermion contribution to the effective action thus is given by
\begin{equation}\label{}
 S_{\rm eff}^{\rm (ferm)}=-\frac{1}{2}\,\mathop{\mathrm{Sp'_{1\oplus 3}}}
 \begin{pmatrix}
   \bar{D}  & \mathop{\mathrm{ad}}K  \\
     \mathop{\mathrm{ad}} \bar{K} & D  \\
 \end{pmatrix}.
\end{equation}

Expanding to the second order in $\mathop{\mathrm{ad}}K$,
$\mathop{\mathrm{ad}}\bar{K}$, we find:
\begin{eqnarray}\label{}
&&\frac{1}{4}\int_{}^{}\frac{d^2p}{\left(2\pi
\right)^2}\,\,\frac{1}{p^2}
 \int_{}^{}d^2x\,
 \left(\,{\rm tr}'_1\mathop{\mathrm{ad}}K\mathop{\mathrm{ad}}\bar{K}
 +\,{\rm tr}'_3\mathop{\mathrm{ad}}\bar{K}\mathop{\mathrm{ad}}K\right)
 \nonumber \\ \nonumber
 &&=\frac{1}{8\pi }\,\ln\Lambda
\int_{}^{}d^2x\,
 \left(\,{\rm tr}_1\mathop{\mathrm{ad}}K\mathop{\mathrm{ad}}\bar{K}
 +\,{\rm tr}_3\mathop{\mathrm{ad}}\bar{K}\mathop{\mathrm{ad}}K\right)
\end{eqnarray}
The prime in the trace is omitted in the second line because the
integrand is proportional to the kappa-symmetry projector. Now,
\begin{eqnarray}\label{}
 \mathop{\mathrm{ad}}K\mathop{\mathrm{ad}}\bar{K}&=&\mathop{\mathrm{ad}}K_\mu
 \mathop{\mathrm{ad}}K^\mu -i\varepsilon ^{\mu \nu
 }\mathop{\mathrm{ad}}K_\mu \mathop{\mathrm{ad}}K_\nu
 \nonumber \\ \nonumber
 \mathop{\mathrm{ad}}\bar{K}\mathop{\mathrm{ad}}{K}&=&\mathop{\mathrm{ad}}K_\mu
 \mathop{\mathrm{ad}}K^\mu +i\varepsilon ^{\mu \nu
 }\mathop{\mathrm{ad}}K_\mu \mathop{\mathrm{ad}}K_\nu,
\end{eqnarray}
so it might seem that fermions renormalize also the operator $\Sigma
_{ab}\varepsilon ^{\mu \nu }K_\mu ^aK_\nu ^b$, where $\Sigma _{ab}$
is an anti-symmetric invariant tensor on $\mathfrak{h}_2$. However,
this operator is a total derivative, its variation being
proportional to $\varepsilon ^{\mu \nu }D_\mu K_\nu =0$, and
integrates to zero.

Adding together bosonic and fermionic contributions we find:
\begin{equation}\label{effs}
 S_{\rm eff} = \frac{1}{8\pi }\,\ln\Lambda
 \int_{}^{}d^2x\,\,\left(2\,{\rm tr}_2-{\rm tr}_1-
 {\rm tr}_3\right)\left(\mathop{\mathrm{ad}}K_\mu
 \right)^2+{\rm finite}.
\end{equation}
Finally, recalling that $K_\mu \in\mathfrak{h}_2$ and thus
$\mathop{\mathrm{ad}}K_\mu $ maps $\mathfrak{h}_2$ to
$\mathfrak{h}_0$ and vice versa, we find:
$$
 \,{\rm tr}_2\mathop{\mathrm{ad}}K_\mu \mathop{\mathrm{ad}}K^\mu =
 \,{\rm tr}_0\mathop{\mathrm{ad}}K^\mu \mathop{\mathrm{ad}}K_\mu.
$$
Hence we can replace $2\,{\rm tr}_2-{\rm tr}_1-
 {\rm tr}_3$ in (\ref{effs}) by ${\rm tr}_0+{\rm tr}_2-{\rm tr}_1-
 {\rm tr}_3=\mathop{\mathrm{Str}}_{\rm adj}$.

If we denote the Hermitian generators of $\mathfrak{h}_2$  by $T_a$
and introduce the metric on the bosonic section of the coset:
\begin{equation}\label{metcos}
g_{ab}=\frac{1}{\kappa ^2}\,\mathop{\mathrm{Str}}T_aT_b,
\end{equation}
the one loop beta-function is
\begin{equation}\label{}
 \beta _{ab}^{\rm 1-loop}=\frac{d}{d\ln\Lambda }\,g_{ab}
 =-\frac{1}{4\pi }\,f^A_{aB}f^B_{bA}(-1)^{|A|},
\end{equation}
where $f^A_{BC}$ are the structure constants of $\mathfrak{g}$. The
beta-function is thus proportional to the Killing form.  The same
one-loop beta-function arises in the pure-spinor-type cosets
\cite{Berkovits:1999zq,Kagan:2005wt,Adam:2007ws}, the supergroup
principal field \cite{Berkovits:1999im,Bershadsky:1999hk}, and in
the $\mathbbm{Z}_2$ cosets of supergroups
\cite{Read:2001pz,Babichenko:2006uc}. The condition for the one-loop
beta-function to vanish is that the Killing form of $\mathfrak{g}$
vanishes\footnote{Strictly speaking, only the projection of the
Killing form on $\mathfrak{h}_2$ should vanish, but if the Killing
form is non-vanishing it is also non-degenerate and unique
\cite{Kac:1977qb,Frappat:1996pb}, and consequently proportional to
the tree-level action thus giving a non-zero beta-function.}.

The calculation of the central charge for the Green-Schwarz string
requires careful regularization of the integration measure
\cite{Kraemmer:1989af,Bastianelli:1990xn,Porrati:1991ts,Bellucci:1991hy},
and yields the following result
\cite{Carlip:1986cz,Kallosh:1988wv,Wiegmann:1989md,Polyakov:2004br}:
the bosons have central charge 1; the left (right) moving fermions
contribute 2 to the left (right) central charge. In our case, $X_3$
and $X_1$ are, respectively, left and right movers so, in total,
\begin{equation}\label{centralleftright}
 c_L=\dim\mathfrak{h}_2+2(\dim\mathfrak{h}_3-N_\kappa),\qquad
 c_R=\dim\mathfrak{h}_2+2(\dim\mathfrak{h}_1-N_{\tilde{\kappa }}).
\end{equation}
The average central charge, $c=(c_L+c_R)/2$,  is determined by the
dimension of the coset and the full rank of the kappa-symmetry:
\begin{equation}\label{mcent}
 c=\dim G/H_0 -N_\kappa -N_{\tilde{\kappa }}.
\end{equation}
The central charge is manifestly positive, in contradistinction to
the non-unitary $\mathbbm{Z}_2$ supercosets, which can have negative
central charge \cite{Babichenko:2006uc} \footnote{Typically, the
central charge of a $\mathbbm{Z}_2$ coset is equal to its
superdimension \cite{Ashok:2009xx} which counts bosons and fermions
with opposite signs.}. By an explicit calculation we will find that
in all conformal cosets $c_L=c_R$. We can thus make no distinction
between $c$, $c_L$ and $c_R$.

We will be also interested in the case when an external CFT is added
to the coset. At first sight, this cannot change the central charge
counting, because the coset and the external CFT interact only via
2d metric which does not carry dynamical degrees of freedom and can
be eliminated by fixing the conformal gauge. However, this is not
quite true. Adding an external CFT can partially or completely break
the kappa-symmetry. The kappa-symmetry transformations act on the 2d
metric and since the latter enters the action of the external CFT,
kappa-symmetry gets broken. In the conformal gauge, the
kappa-symmetry breaking can be attributed to the violation of the
null condition for the currents (\ref{null}), which does not hold in
the presence of another CFT with a non-trivial energy-momentum
tensor\footnote{The observation that coupling to an external CFT
breaks kappa-symmetry was made in \cite{Babichenko:2009dk}, but
perhaps this simple fact was known before.}.

The ranks of the left- and right-moving kappa-symmetries with the
null condition relaxed will be denoted by $\hat{N}_\kappa $,
$\hat{N}_{\tilde{\kappa }}$. They are computed by the same formulas
(\ref{rankkappa}) where $K$ and $\bar{K}$ are now the most general
elements of $\mathfrak{h}_2$, not necessarily null. We will denote
the central charge of the sigma-model coupled to an external CFT
(the extrinsic central charge) by\footnote{This should not be
confused with $2c/3$ sometimes also denoted by $\hat{c}$.}
$\hat{c}$:
\begin{equation}\label{hatcent}
 \hat{c}=\dim G/H_0 -\hat{N}_\kappa -\hat{N}_{\tilde{\kappa }},
\end{equation}
In the next section we will compute extrinsic and intrinsic central
charges for all conformal $\mathbbm{Z}_4$ cosets.

%%%%%%%%%%%%%%%%%%%%%%%%%%%%%%%%%%%%%%%%%%%%%%%%%%%%%%%%%%%%%%%%%%%%%%%%%%%%%%%%
\section{Conformal Sigma Models}\label{conforms}

The string sigma-model must be defined on a real superspace, so the
symmetry algebra $\mathfrak{g}$ should be a real Lie superalgebra.
However, the one-loop beta-function and the central charge depend
only on the structure constants of $\mathfrak{g}$ and therefore are
the same for all real forms of a given complex superalgebra. Dealing
with complex Lie superalgebras is technically simpler, and
subsequent analysis will be done as if $\mathfrak{g}$ were complex.
We will pick a particular real form in the very end. If we want to
have a string interpretation of the sigma-model, the real form must
be such that the metric (\ref{metcos}) has the Minkowski signature
$(-+\ldots +)$. In the cases when the requisite real form does not
exist, we will keep in  mind the compact form of the coset with the
$(+\ldots +)$ metric.

The basic complex Lie superalgebras with vanishing Killing form form
two infinite series: $\mathfrak{psu}(n|n)$ and
$\mathfrak{osp}(2n+2|2n)$ \cite{Kac:1977qb,Frappat:1996pb}. The
one-parameter family of exceptional superalgebras
$\mathfrak{d}(2,1;\alpha )$, a continuous deformation of
$\mathfrak{osp}(4|2)$,   also has vanishing Killing form. But since
the deformation parameter appears only in the anti-commutator of
supercharges, the central charge counting for
$\mathfrak{d}(2,1;\alpha )$ is the same as for $\mathfrak{osp}(4|2)$
and we need not discuss $\mathfrak{d}(2,1;\alpha )$ separately, just
keeping in mind that any $OSp(4|2)$ coset can be generalized to
$D(2,1;\alpha )$.

From the discussion above we see that there are two series of
conformal sigma-models on semi-symmetric superspaces, those with
$PSU(n|n)$ and $OSp(2n+2|2n)$ symmetry, which we will call type-$U$
and type-$O$ models. All possible $\mathbbm{Z}_4$ automorphisms of
$\mathfrak{psu}(n|n)$ and $\mathfrak{osp}(2n+2|2n)$ and the
corresponding cosets were classified by Serganova \cite{Serganova}.
They fall into six separate classes, four type-$U$ and two type-$O$,
conveniently described with the help of the supermatrix
representation of the $\mathfrak{su}(n|n)$ and
$\mathfrak{osp}(2n+2|2n)$ superalgebras\footnote{The central element
in $\mathfrak{su}(n|n)$ that distinguishes it from
$\mathfrak{psu}(n|n)$ can be trivially factored out.}:
\begin{eqnarray}\label{sunnmatrix}
 \mathfrak{su}(n|n)&=&\left\{X\in M(n|n)\left| \right.\mathop{\mathrm{Str}}X=0
\right\}
 \\ \label{osp2n2nmatrix}
 \mathfrak{osp}(2n+2|2n)&=&\left\{
 \begin{pmatrix}
    A & \Theta   \\
    \Psi  & B  \\
 \end{pmatrix}\in M(2n+2|2n)\left|
 \vphantom{\begin{pmatrix}
    A & \Theta   \\
    \Psi  & B  \\
 \end{pmatrix}}
 \right.
 A=-A^t, B=JB^tJ, \Psi =J\Theta ^t
 \right\}.
\end{eqnarray}
Here $J$ is the $2n\times 2n$ matrix
\begin{equation}\label{}
 J=
 \begin{pmatrix}
   0  & \mathbbm{1}_{n\times n}  \\
    -\mathbbm{1}_{n\times n} & 0  \\
 \end{pmatrix}.
\end{equation}
We will also need the diagonal matrix
\begin{equation}\label{whatsip}
I_p=\mathop{\mathrm{diag}}\left(\mathbbm{1}_{p\times p},
-\mathbbm{1}_{(n-p)\times (n-p)} \right),
\end{equation}
and the following supermatrix operations\footnote{Our notations are
essentially identical to those of \cite{Serganova}.}:
\begin{eqnarray}\label{opera}
  \begin{pmatrix}
   A  & \Theta   \\
   \Psi   &  B \\
 \end{pmatrix}^{st}&=&
 \begin{pmatrix}
   A^t  & -\Psi ^t  \\
    \Theta ^t & B^t  \\
 \end{pmatrix}
 \nonumber \\
  \delta\circ\begin{pmatrix}
   A  & \Theta   \\
   \Psi   &  B \\
 \end{pmatrix}&=&
 \begin{pmatrix}
   A  & i\Theta   \\
   -i\Psi   & B  \\
 \end{pmatrix}
 \nonumber \\
 \Pi\circ\begin{pmatrix}
   A  & \Theta   \\
   \Psi   &  B \\
 \end{pmatrix}
 &=&
 \begin{pmatrix}
   B  & \Psi   \\
   \Theta   & A  \\
 \end{pmatrix}
\end{eqnarray}
These three operations and the adjoint action of the matrices $J$
and $I_p$ allow one to build all possible $\mathbbm{Z}_4$
automorphisms of $\mathfrak{su}(n|n)$  (table~\ref{tabu})
\begin{table}[t]
\caption{\small Semi-symmetric cosets of $PSU(n|n)$. $\Omega $ is
the $\mathbbm{Z}_4$ automorphism of $\mathfrak{su}(n|n)$, $H_0$ is
the invariant subgroup. The bosonic section of the coset is
$SU(n)\times SU(n)/H_0$.}\label{tabu}\centering
\begin{tabular}{|c|c|c|}
\hline
 {\bf Coset} &  $\mathbf{\Omega} $ & $\mathbf{H_0}$ \\
 \hline
 type-$U1$ & $\mathop{\mathrm{Ad}}\mathop{\mathrm{diag}}(I_p,I_q)\circ\delta $ &
 $U(p)\times SU(n-p)\times U(q)\times SU(n-q)$ \\
 \hline
 type-$U2$ & $-st$ & $SO(n)\times SO(n)$ \\
 \hline
 type-$U3$ & $ -st\circ\Pi\circ\delta $ & $SU(n)$ \\
 \hline
 type-$U4$ &
 $-st\circ\mathop{\mathrm{Ad}}\mathop{\mathrm{diag}}(J,J)$ & $Sp(n)\times
 Sp(n)$ \\
 \hline
\end{tabular}
\end{table}
and $\mathfrak{osp}(2n+2|2n)$ (table~\ref{tabo}) \cite{Serganova}.
\begin{table}[t]
\caption{\small Semi-symmetric cosets of $OSp(2n+2|2n)$. $\Omega $
is the $\mathbbm{Z}_4$ automorphism of $\mathfrak{osp}(2n+2|2n)$,
$H_0$ is the invariant subgroup. The bosonic section of the coset is
$SO(2n+2)\times Sp(2n)/H_0$.}\label{tabo}\centering
\begin{tabular}{|c|c|c|}
\hline
 {\bf Coset} & $\mathbf{\Omega} $ & $\mathbf{H_0}$ \\
 \hline
 type-$O1$ & $\mathop{\mathrm{Ad}}\mathop{\mathrm{diag}}(I_p,J)$ &
 $SO(p)\times SO(2n+2-p)\times U(n)$ \\
 \hline
 type-$O2$ &
 $\mathop{\mathrm{Ad}}\mathop{\mathrm{diag}}(J,\mathbbm{1}\otimes I_p)$ &
 $U(n+1)\times
 Sp(2p)\times Sp(2n-2p)$ \\
 \hline
\end{tabular}
\end{table}
The $\mathbbm{Z}_4$ decomposition corresponding to these cosets is
described in more detail in the appendix.

Almost all of the cosets in tables~\ref{tabu} and \ref{tabo} can be
used to define the action of a sigma-model, except for type-$U3$.
Any element of $\mathfrak{h}_2$ in a type-$U3$ coset is null, at
least in the usual supertrace metric (the explicit $\mathbbm{Z}_4$
decomposition is given in sec.~\ref{typu3}). The bosonic part of the
action then vanishes identically. Although we will formally compute
the central charge for this coset, we will not discuss this
sigma-model any further.

In addition to the models based on simple Lie superalgebras one can
also consider the cosets of product groups. Such cosets naturally
arise in the $AdS_3/CFT_2$ correspondence, because the conformal
algebra in two dimensions is a direct sum of two Virasoro algebras
acting independently on the left and right movers. Independently of
the $AdS/CFT$ connection, the product structure is quite natural
from the point of view of the coset construction, as it generally
admits a $\mathbbm{Z}_4$ action. If $\mathfrak{p}$ is a
superalgebra, we can define a $\mathbbm{Z}_4$ action on the direct
sum $\mathfrak{g}=\mathfrak{p}\oplus\mathfrak{p}$ by combining the
permutation of the two factors with the fermion number
\cite{Babichenko:2009dk}:
\begin{equation}\label{omegad}
 \Omega =
 \begin{pmatrix}
   0  & \mathop{\mathrm{id}}  \\
    (-1)^F & 0  \\
 \end{pmatrix}.
\end{equation}
One can easily check that $\Omega ([X,Y])=[\Omega (X),\Omega (Y)]$
for any $X,Y\in\mathfrak{p}\oplus\mathfrak{p}$. It is also obvious
that $\Omega ^2=(-1)^F$ and thus $\Omega ^4=\mathop{\mathrm{id}}$.
The invariant subalgebra of the $\mathbbm{Z}_4$ action is the
bosonic diagonal $\mathfrak{h}_0=\{(X,X)|X\in\mathfrak{p}\}$.
Consequently, the supercoset is $P\times P/H_0$, where $H_0$ is the
bosonic subgroup of $P$ diagonally embedded in $P\times P$. The
bosonic section is the group manifold of $H_0$. We refer to the
tensor-product semi-symmetric spaces as type-$Tu$ cosets if
$P=PSU(n|n)$ and type-$To$ cosets if $P=OSp(2n+2|2n)$. There are
also interesting cosets of $U(n|n)$ \cite{Stefanski:2007dp}, which
we will not consider here.

We will calculate the one-loop central charge for the eight types of
semi-symmetric cosets introduced above (type-$U1$-$4$, type-$O1$,$2$
and type-$Tu$,$o$). The central charge counts the number of degrees
of freedom in the sigma-model and depends on the rank of the
kappa-symmetry (\ref{centralleftright})--(\ref{hatcent}), which in
turn is given by the dimension of the commutant of a generic element
of $\mathfrak{h}_2$ (\ref{rankkappa}). The calculations reduce to
simple algebra, but have to be done case by case. The details are
given in the appendix, here we just describe the general pattern
that emerges:
\begin{itemize}
\item
The left- and right-moving kappa-symmetries, which are associated
with the $\mathfrak{h}_3$ and $\mathfrak{h}_1$ subspaces,  are
identical in almost all the cases. One and only exception is the
type-$U3$ coset, for  which  $\mathfrak{h}_3$ and $\mathfrak{h}_1$
are not isomorphic and have different dimensions. The kappa-symmetry
compensates for this, such that even in this case the left- and
right-moving central charges are equal.
\item
The extrinsic kappa-symmetries follow a regular pattern and depend
uniformly on the dimensionalities of the superalgebra and the coset
(table~\ref{kaprankgen})\footnote{For type-$U1$, we assume that
$n\geq 2p$ and $n\geq 2q$.}.
\begin{table}[t]
\caption{\small The rank of the kappa-symmmetry, generic case.}
\label{kaprankgen}\centering
\begin{tabular}{|c|c|c|}
\hline
 {\bf Coset} & $\mathbf{\hat{N}_\kappa } $ & $\mathbf{\hat{N}_{\tilde{\kappa }}}$ \\
 \hline
 type-$U1$ & $(n-2p)(n-2q)$ & $(n-2p)(n-2q)$
  \\
 \hline
 type-$U2$ & $0$ & $0$  \\
 \hline
 type-$U3$ & $0$ & $2n$ \\
 \hline
 type-$U4$ &
 $0$ & $0$ \\
 \hline
 type-$O1$ & $0$ &
 $0$ \\
 \hline
 type-$O2$ ($n~{\rm odd}$) &
 $0$ &
 $0$ \\
 \hline
 type-$O2$ ($n~{\rm even}$) &
 $2n-4p$ &
 $2n-4p$ \\
 \hline
 type-$Tu$ &
 $0$ & $0$ \\
 \hline
 type-$To$ &
 $0$ & $0$ \\
 \hline
\end{tabular}
\end{table}
\item
There is no difference between intrinsic and extrinsic
kappa-sym\-met\-ries and central charges in most cases, but in low
ranks there are exceptions  listed in table~\ref{exkaprank}.
Imposing the Virasoro constraints then increases the rank of the
kappa-symmetry and decreases the central charge.
\begin{table}[t]
\caption{\small The rank of the kappa-symmetry, exceptions (in all
cases $N_{\tilde{\kappa }}=N_\kappa $).} \label{exkaprank}\centering
\begin{tabular}{|c|c|}
\hline
 {\bf Coset} & $\mathbf{{N}_\kappa } $  \\
 \hline
 type-$U1$ ($p=1, q=1$)  & $n^2-4n+6$
  \\
 \hline
 type-$U2$ ($n=2$) & $2$   \\
 \hline
 type-$U4$ ($n=4$) &
 $8$  \\
 \hline
 type-$O1$ ($n=1, p=1$) & $1$ \\
 \hline
 type-$O2$ ($n=2, p=1$) &
 $4$ \\
 \hline
 type-$Tu$ ($n=2$) &
 $4$  \\
 \hline
\end{tabular}
\end{table}
\end{itemize}
The central charges for the regular cosets (for which there is no
difference between $\hat{c}$ and $c$) are summarized in
table~\ref{centrgen}. The exceptional cases in which the intrinsic
and extrinsic central charges are different are listed in
table~\ref{excentr}.
\begin{table}[t]
\caption{\small The central charge, regular case.}
\label{centrgen}\centering
\begin{tabular}{|c|c|}
\hline
 {\bf Coset} & $\mathbf{\hat{c}} $  \\
 \hline
 type-$U1$ & $6(p+q)n-2p^2-2q^2-8pq$
  \\
 \hline
 type-$U2$ & $3n^2+n-2$  \\
 \hline
 type-$U3$ & $3n^2-2n-1$ \\
 \hline
 type-$U4$ &
 $3n^2-n-2$  \\
 \hline
 type-$O1$ &
 $5n^2+(2p+5)n-p^2+2p$ \\
 \hline
 type-$O2$ ($n~{\rm odd}$) &
 $5n^2+(4p+5)n-4p^2$  \\
 \hline
 type-$O2$ ($n~{\rm even}$) &
 $5n^2+(4p+1)n-4p^2+8p$ \\
 \hline
 type-$Tu$ &
 $6n^2-2$ \\
 \hline
 type-$To$ &
 $12n^2+12n+1$  \\
 \hline
\end{tabular}
\end{table}
\begin{table}[t]
\caption{\small The central charge, exceptions.}
\label{excentr}\centering
\begin{tabular}{|c|c|c|}
\hline
 {\bf Coset} & $\mathbf{c} $ & $\mathbf{\hat{c}}$ \\
 \hline
 type-$U1$ ($p=1, q=1$)  & $12n-16$ & $12n-12$
  \\
 \hline
 type-$U2$ ($n=2$) & $8$ & $12$  \\
 \hline
 type-$U4$ ($n=4$) &
 $26$ & $42$ \\
 \hline
 type-$O1$ ($n=1, p=1$) & $11$ & $13$ \\
 \hline
 type-$O2$ ($n=2, p=1$) &
 $26$ & $34$ \\
 \hline
 type-$Tu$ ($n=2$) &
 $14$ & $22$ \\
 \hline
\end{tabular}
\end{table}

%%%%%%%%%%%%%%%%%%%%%%%%%%%%%%%%%%%%%%%%%%%%%%%%%%%%%%%%%%%%%%%%%%%%%%%%%%%%%%%%
\section{String sigma-models}

The worldsheet diffeomorphisms in the sigma-models at hand are just
the same as in the bosonic string theory, and lead to the same set
of $bc$ ghosts in the conformal gauge. If the ghost contribution to
the central charge is to be canceled by the sigma-model alone, its
intrinsic central charge should be equal to $26$.  If $\hat{c}<26$,
the central charge deficit can be compensated by coupling to an
external CFT with central charge $26-\hat{c}$. Let us list the
models that satisfy these criteria.

The inspection of tables~\ref{centrgen} and \ref{excentr} shows that
there are only two cosets with $c=26$, both are exceptional:
\begin{eqnarray}\label{criticalstrings}
&&{\rm Type-}U4 (n=4): \qquad   PU(2,2|4)/SO(4,1)\times SO(5) \qquad
 AdS_5\times S^5
 \nonumber \\
&&{\rm Type-}O2 (n=2,p=1): \qquad  OSp(6|4)/U(3)\times SO(3,1)
\qquad
 AdS_4\times CP^3
\end{eqnarray}
These cosets define well-known sigma-models. The first one is the
Metsaev-Tseytlin model for the Green-Schwarz superstring on
$AdS_5\times S^5$ \cite{Metsaev:1998it}. The second model describes
strings on $AdS_3\times CP^3$
\cite{Arutyunov:2008if,Stefanski:2008ik} and can be obtained from
the Green-Schwarz action on this background \cite{Gomis:2008jt} by
partially fixing kappa-symmetry. It is interesting that these two
cases are in a sense unique.

There is a number of non-critical semi-symmetric cosets with
$\hat{c}<26$. Some of them admit a real form with the Minkowski
metric on the bosonic subspace:
\begin{eqnarray}\label{noncriticalstrings}
 &&1)~{\rm Type-}To (n=1): \qquad OSp(4|2)\times OSp(4|2)/SO(4)\times
 SL(2,\mathbbm{R}) \nonumber \\
 &&\qquad \hat{c}=25 \qquad  AdS_3\times S^3\times S^3
 \nonumber\\
 &&2)~{\rm Type-}Tu (n=2):\qquad PSU(1,1|2)\times PSU(1,1|2)/SU(1,1)\times
 SU(2) \nonumber \\
 &&\qquad \hat{c}=22\qquad AdS_3\times S^3
 \nonumber \\
 &&3)~{\rm Type-}O1 (n=1,p=2):\qquad OSp(4|2)/U(1)^3\nonumber \\
 &&\qquad \hat{c}=14\qquad AdS_2\times S^2\times S^2
 \nonumber \\
 &&4)~{\rm Type-}O1 (n=1,p=1):\qquad OSp(4|2)/SO(3)\times U(1)\nonumber \\
 &&\qquad \hat{c}=13\qquad AdS_2\times S^3
 \nonumber \\
 &&5)~{\rm Type-}U1 (n=2,p=1,q=1)/U2 (n=2):\qquad PSU(1,1|2)/U(1)^2\nonumber \\
 &&\qquad \hat{c}=12\qquad AdS_2\times S^2
 \nonumber \\
 &&6)~{\rm Type-}U1 (n=2,p=1,q=0):\qquad PSU(1,1|2)/U(1)\times SU(2)\nonumber \\
 &&\qquad \hat{c}=10\qquad AdS_2
 \nonumber \\
 &&7)~{\rm Type-}O1 (n=1,p=0)/O2 (n=1,p=0):\qquad OSp(4|2)/SO(4)\times
 U(1)\nonumber \\
 &&\qquad \hat{c}=10\qquad AdS_2
\end{eqnarray}
Here $OSp(4|2)$ can be replaced by a more general supergroup
$D(2,1;\alpha )$, which gives a one-parametric family of
sigma-models. In particular, there is a $D(2,1;\alpha )$ coset which
continuously interpolates between cases 7 and 6. The same is true
for cases 1 and 2 \cite{Babichenko:2009dk}, where the degeneration
of $D(2,1;\alpha )$ to $PSU(1,1|2)$ leaves two extra flat dimensions
which account for the difference in central charges.

Many of the cosets above  have been discussed in the context of the
AdS/CFT duality. The first coset, supplemented by an external $S^1$,
describes the Green-Schwarz string on $AdS_3\times S^3\times
S^3\times S^1$ with completely fixed kappa-symmetry
\cite{Babichenko:2009dk}. The action of the second coset can be
interpreted as the 6d Green-Schwarz action on $AdS_3\times S^3$
\cite{Rahmfeld:1998zn,Pesando:1998wm,Park:1998un,Metsaev:2000mv,Chen:2005uj,Adam:2007ws}
 and as such admits rank-eight kappa-symmetry
(table~\ref{exkaprank}). However, coupling to an external $T^4$,
which is necessary to compensate for the central charge deficit,
breaks kappa-symmetry and changes the central charge counting. This
coset plus four compact bosons describes the Green-Schwarz string
on\footnote{This background admits a hybrid description in terms of
the sigma model on the supergroup manifold $PSU(1,1|2)$
\cite{Berkovits:1999im}.} $AdS_3\times S^3\times T^4$ with fully
fixed kappa-symmetry \cite{Babichenko:2009dk}. The fifth coset
yields the 4d Green-Schwarz action on $AdS_2\times S^2$
\cite{Zhou:1999sm,Berkovits:1999zq,Adam:2007ws}. Again, its
(four-parameter) kappa-symmetry is completely broken by coupling to
an external $c=14$ CFT. The models 3, 4, 6 and 7 are seemingly new.
The last two models are similar to the $OSp(1|2)/U(1)$ coset
considered in \cite{Verlinde:2004gt} -- they have $AdS_2$ as the
bosonic target space and no physical degrees of freedom on shell.
The latter is due to kappa-symmetry. Coupling of these sigma-models
to an external CFT breaks kappa-symmetry and revives their fermion
degrees of freedom.

Other non-critical cosets do not admit a metric with the $(-+\ldots
+)$ signature and the time direction should lie in the external CFT.
Their compact versions (those with the Euclidean metric) are listed
below:
\begin{eqnarray}\label{compact}
 &&1)~{\rm Type-}U1 (n=3,p=1,q=1):\qquad PSU(1,2|3)/U(2)\times U(2)\nonumber \\
 &&\qquad \hat{c}=24\qquad CP^2\times CP^2
 \nonumber \\
 &&2)~{\rm Type-}U1 (n=4,p=1,q=0):\qquad PSU(4|4)/U(3)\times SU(4)\nonumber \\
 &&\qquad \hat{c}=22\qquad CP^3
 \nonumber \\
 &&3)~{\rm Type-}O2 (n=2,p=0):\qquad OSp(6|4)/U(3)\times Sp(4)\nonumber \\
 &&\qquad \hat{c}=22\qquad CP^3
 \nonumber \\
 &&4)~{\rm Type-}U1 (n=3,p=1,q=0):\qquad PSU(3|3)/U(2)\times SU(3)\nonumber \\
 &&\qquad \hat{c}=16\qquad CP^2.
\end{eqnarray}
These models bear certain resemblance to the $\mathbbm{CP}^{S-1|S}$
$\mathbbm{Z}_2$ sigma models \cite{Candu:2009ep}.

%%%%%%%%%%%%%%%%%%%%%%%%%%%%%%%%%%%%%%%%%%%%%%%%%%%%%%%%%%%%%%%%%%%%%%%%%%%%%%%%
\section{Conclusions}

The list of semi-symmetric superspaces potentially consistent as
string backgrounds is not very long. We should stress that we have
computed the beta function and central charge only at the one loop
level. There is no guarantee that higher-order corrections
identically vanish, and the list of consistent string backgrounds
with the $\mathbbm{Z}_4$ symmetry may be even shorter. It is
instructive to look at what happens in the principal chiral models
and $\mathbbm{Z}_2$ cosets of supergroups. In the case of the
principal chiral field, it is possible to prove finiteness to all
orders in perturbation theory for the cosets with the vanishing
one-loop beta-function \cite{Bershadsky:1999hk,Babichenko:2006uc}.
Many one-loop finite $\mathbbm{Z}_2$  cosets are two-loop finite as
well \cite{Babichenko:2006uc}, but the full set of conformal
$\mathbbm{Z}_2$ cosets seems to be smaller than the set of
$\mathbbm{Z}_2$ cosets with vanishing one-loop beta-function
\cite{Candu:2010yg}.

The semi-symmetric cosets with non-zero beta-function can also be
interesting for the AdS/CFT duality, if they are asymptotically
free. A sigma-model with the AdS target cannot develop a mass gap
because of the non-compactness. The asymptotic freedom at weak
coupling then suggests that the beta-function has a non-trivial
zero, which can potentially be interpreted as string theory on the
AdS space of fixed radius \cite{Polyakov:2004br}\footnote{This
argument was suggested to the author by A.M.~Polyakov.}.

The consistent Minkowski backgrounds, critical
(\ref{criticalstrings}) and non-critical (\ref{noncriticalstrings}),
all involve an AdS factor and are potentially dual to CFTs in
dimensions $d\leq 4$. In all these cases the worldsheet sigma-model
is integrable and thus potentially solvable by Bethe ansatz. For the
string sigma-models on $AdS_5\times S^5$ and $AdS_4\times CP^3$ the
classical algebraic curve
\cite{Kazakov:2004qf,Beisert:2005bm,Gromov:2008bz}, the worldsheet
S-matrix \cite{Beisert:2005tm,Ahn:2008aa} and the asymptotic quantum
Bethe equations \cite{Beisert:2005fw,Gromov:2008qe} are known. The
finite-volume TBA/Y-system solution is now also available
\cite{Gromov:2009tv,Bombardelli:2009ns,Gromov:2009bc,Arutyunov:2009ur,Bombardelli:2009xz,Gromov:2009at}.
It would be interesting to derive a unifying Bethe-ansatz solution
for a generic semi-symmetric coset.

%%%%%%%%%%%%%%%%%%%%%%%%%%%%%%%%
\subsection*{Acknowledgments}
I would like to thank I.~Adam, A.~Babichenko, I.~Bakas,
N.~Berkovits, V.~Kazakov, S.~Leurent, L.~Mazzucato, J.~Minahan,
V.~Mitev, Y.~Oz, A.~Polyakov, V.~Schomerus, S.~Sha\-ta\-shvi\-li,
B.~Stefanski, A.~Tseytlin, J.~Troost, B.~Vicedo, K.~Wiese and
N.~Wyllard for interesting discussions and useful comments. This
work was supported in part by the BQR ENS, in part by the Swedish
Research Council under the contract 621-2007-4177, in part by the
ANF-a grant 09-02-91005, and in part by the grant for support of
scientific schools NSH-3036.2008.2.
%%%%%%%%%%%%%%%%%%%%%%%%%%%%%%%%%%

\appendix

%%%%%%%%%%%%%%%%%%%%%%%%%%%%%%%%%%%%%%%%%%%%%%%%%%%%%%%%%%%%%%%%%%%%%%%%%%%%%%%%
\section{Rank of kappa symmetry}\label{app1}

In this appendix we compute the rank of the kappa-symmetry for all
conformal $\mathbbm{Z}_4$ cosets. According to (\ref{rankkappa}),
the rank is equal to the dimension of the commutant of a generic
element $K({\rm or}~\bar{K})\in\mathfrak{h}_2$ in $\mathfrak{h}_1$
and $\mathfrak{h}_3$. In the supermatrix representation,
\begin{equation}\label{}
 K({\rm or~}\bar{K})=
 \begin{pmatrix}
   A  & 0  \\
   0  & B  \\
 \end{pmatrix}.
\end{equation}
Commuting this with an odd element of the superalgebra, we find:
$$
 \left[
 \begin{pmatrix}
   A  &  0 \\
   0  &  B \\
 \end{pmatrix},
 \begin{pmatrix}
   0  & \Theta   \\
   \Psi   & 0  \\
 \end{pmatrix}
 \right]=
 \begin{pmatrix}
   0  & A\Theta -\Theta B  \\
   B\Psi-\Psi A   &  0 \\
 \end{pmatrix}.
$$
The commutator vanishes if
\begin{equation}\label{comma}
 A\Theta =\Theta B,\qquad B\Psi =\Psi A.
\end{equation}
The number of solutions to these equations determines the rank of
the kappa-symmetry. The dimension of the solution space for generic
$A$ and $B$ determines the rank of the extrinsic kappa-symmetry
$\hat{N}_\kappa $, $\hat{N}_{\tilde{\kappa }}$. To compute the rank
of the intrinsic kappa-symmetries $N_\kappa $, $N_{\tilde{\kappa
}}$, one should in addition impose the Virasoro constraints
(\ref{null}). The matrices $A$ and $B$ then satisfy
\begin{equation}\label{virnull}
 \,{\rm tr}\,A^2=\,{\rm tr}\,B^2.
\end{equation}
Throughout the calculation we will use a number of simple algebraic
facts, which we collect below.

Consider an equation for an $m\times n$ matrix $X$:
\begin{equation}\label{mateq}
 MX=XN,
\end{equation}
where $M$ and $N$ are given quadratic matrices, which we assume to
be sufficiently generic. Since a generic matrix can be diagonalized
by a similarity transformation, without loss of generality we can
assume that $M$ and $N$ are diagonal. Denoting their eigenvalues by
$\mu _i$, $i=1,\ldots ,m$ and $\nu _a$, $a=1,\ldots ,n$, we find
\begin{equation}\label{}
 (\mu _i-\nu _a)X_{ia}=0.
\end{equation}
In the most general case of arbitrary $M$ and $N$, all $\mu _i$ are
different from $\nu _a$, and consequently (\ref{mateq}) has no
solutions other than $X=0$. This might not be true if $M$ and $N$
satisfy extra conditions. For instance, if $M$ and $N$ are $2\times
2$ matrices constrained by
\begin{equation}\label{2by2}
 \,{\rm tr}\,M=0=\,{\rm tr}\,N,\qquad \,{\rm tr}\,M^2=\,{\rm
 tr}\,N^2,
\end{equation}
their eigenvalues coincide pairwise: $\mu _1=-\mu _2=\nu _1=-\nu
_2$. In this case the equation (\ref{mateq}) has two linearly
independent solutions. In general the rank of the linear system
(\ref{mateq}) is equal to the number of pairs of coinciding
eigenvalues of matrices $M$ and $N$.

In analyzing the spectrum of various matrices we will repeatedly use
the Stenzel theorem \cite{Stenzel,Ikramov}, which states that the
non-zero eigenvalues of a product of two anti-symmetric matrices are
doubly degenerate. Namely, the spectrum of an $n\times n$ matrix
$M=A_1A_2$, where $A_i^t=-A_i$, consists of $[n/2]$ pairs of
eigenvalues $\mu _1,\mu _1,\ldots \mu _{[n/2]},\mu _{[n/2]}$ and, if
$n$ is odd, an additional zero eigenvalue associated with the vector
annihilated by $A_2$.

\subsection{Type-$U1$}

The $\mathbbm{Z}_4$ automorphism of the type-$U1$ coset acts on the
supermatrices as
\begin{equation}\label{}
 \Omega \circ
 \begin{pmatrix}
   A  & \Theta   \\
   \Psi   & B  \\
 \end{pmatrix}
 =
 \begin{pmatrix}
   I_pAI_p  & iI_p\Theta I_q  \\
    -iI_q\Psi I_p & I_qBI_q  \\
 \end{pmatrix},
\end{equation}
where $I_p$, $I_q$ are defined in (\ref{whatsip}). The
$\mathbbm{Z}_4$ decomposition in the supermatrix representation is
given by
\begin{eqnarray}\label{}
 \mathfrak{h}_2:&&
 A=
 \begin{pmatrix}
   0  & [A_1]_{p\times (n-p)}  \\
   [A_2]_{(n-p)\times p}  & 0  \\
 \end{pmatrix},\qquad
 B=
\begin{pmatrix}
   0  & [B_1]_{q\times (n-q)}  \\
   [B_2]_{(n-q)\times q}  & 0  \\
 \end{pmatrix}
 \nonumber \\
 \mathfrak{h}_1:&&
 \Theta =
 \begin{pmatrix}
   [\Theta _1]_{p\times q}  & 0  \\
    0 & [\Theta _2]_{(n-p)\times (n-q)}  \\
 \end{pmatrix},\qquad
  \Psi  =
 \begin{pmatrix}
   0  & [\Psi _1]_{q\times (n-p)}  \\
    [\Psi _2]_{(n-q)\times p} & 0  \\
 \end{pmatrix}
 \nonumber \\
 \mathfrak{h}_3:&&
 \Theta =
 \begin{pmatrix}
   0  & [\Theta _1]_{p\times (n-q)}  \\
    [\Theta _2]_{(n-p)\times q} & 0  \\
 \end{pmatrix},\qquad
  \Psi  =
 \begin{pmatrix}
   [\Psi _1]_{q\times p}  & 0  \\
    0 & [\Psi _2]_{(n-q)\times (n-p)}  \\
 \end{pmatrix}
\end{eqnarray}

Let us assume that $n-p\geq p$ and $n-q\geq q$. The non-zero
eigenvalues of the matrix $A$ form $p$ pairs $\pm \alpha _1,\ldots,
\pm\alpha _p$, where $\alpha _i^2$ are the eigenvalues of the
$p\times p$ matrix $A_1A_2$. In addition $A$, has $n-2p$ zero modes
built from $(n-p)$-dimensional vectors $v_i$ annihilated by $A_1$.
Analogously, the right action of $B$ produces $2q$ non-zero
eigenvalues $\pm\beta _1,\ldots ,\pm\beta _q$, whose squares $\beta
_j^2$ are eigenvalues of $B_1B_2$, and $n-2q$ zero modes made of
left $n-q$ dimensional null vectors of $B_2$, which we denote by
$u_j$. There are $(n-2p)(n-2q)$ pairs of zero eigenvalues of $A$ and
$B$. In $\mathfrak{h}_1$, they correspond to $(n-2p)(n-2q)$
solutions to (\ref{comma}) of the form
\begin{equation}\label{}
 \Psi =0,\qquad\Theta _1=0,\qquad \Theta _2=v_i\otimes
 u_j~(i=1,\ldots,
 n-2p;\,j=1,\ldots, n-2q).
\end{equation}
In general, the non-zero eigenvalues $\pm\alpha _i$ have no reasons
to coincide with $\pm\beta _j$, so the number of right-moving
kappa-symmetries without Virasoro constraints is
\begin{equation}\label{}
 \hat{N}_{\tilde{\kappa }}=(n-2p)(n-2q).
\end{equation}
The computation for the left-movers ($\mathfrak{h}_3$) is the same
with the left and right action of $A$ and $B$ interchanged and
$\Theta $ replaced by $\Psi $, so
\begin{equation}\label{}
 \hat{N}_{{\kappa }}=(n-2p)(n-2q).
\end{equation}

The null condition (\ref{virnull}), in terms of the eigenvalues
reads
\begin{equation}\label{sumki}
 \sum_{i=1}^{p}\alpha _i^2=\sum_{j=1}^{q}\beta _j^2.
\end{equation}
In general, this condition is too weak and does not imply any
degeneracies. The only exception is $p=1=q$, when each of the
matrices $A$ and $B$ has only one pair of non-zero eigenvalues. The
eq.~(\ref{sumki}) then implies that these eigenvalues coincide up to
a sign. We thus find two extra solutions to (\ref{comma}) in both
$\mathfrak{h}_1$ and $\mathfrak{h}_3$. In $\mathfrak{h}_1$, the
solutions are
\begin{equation}\label{}
\Theta _1=A_1A_2,\qquad \Theta _2=A_2B_1,\qquad \Psi =0
\end{equation}
and
\begin{equation}\label{}
\Psi _1=A_1,\qquad \Psi _2=B_2,\qquad \Theta =0.
\end{equation}
This solutions exist provided that $A_1A_2=B_1B_2$, which for
$p=1=q$ is a consequence of the Virasoro constraints. Hence,
\begin{equation}\label{}
 N_\kappa =N_{\tilde{\kappa }}=(n-2)^2+2\qquad (p=1=q).
\end{equation}

\subsection{Type-$U2$}

The $\mathbbm{Z}_4$ automorphism of the type-$U2$ coset is
\begin{equation}\label{}
 \Omega \circ
 \begin{pmatrix}
   A  & \Theta   \\
   \Psi   &  B \\
 \end{pmatrix}
 =
 \begin{pmatrix}
   -A^t  & \Psi ^t  \\
    -\Theta ^t & -B^t  \\
 \end{pmatrix},
\end{equation}
which gives the following $\mathbbm{Z}_4$ decomposition:
\begin{eqnarray}\label{}
 \mathfrak{h}_2:&&
 A^t=A,\qquad B^t=B
 \nonumber \\
 \mathfrak{h}_1:&&
 \Psi =i\Theta ^t
 \nonumber \\
 \mathfrak{h}_3:&&
 \Psi =-i\Theta ^t.
\end{eqnarray}

To find the rank of the kappa-symmetry, we need to solve the
equation
\begin{equation}\label{eigenv}
 A\Theta =\Theta B
\end{equation}
for generic symmetric traceless matrices $A$ and $B$. It general it
has no solutions, so
\begin{equation}\label{}
 \hat{N}_\kappa =0=\hat{N}_{\tilde{\kappa }}.
\end{equation}
If we impose the Virasoro condition (\ref{virnull}), we are in the
situation described around eq.~(\ref{2by2}). The exceptional case is
$n=2$, in which the space of solutions to (\ref{eigenv}) is
two-dimensional. We thus find:
\begin{equation}\label{}
 N_\kappa =N_{\tilde{\kappa }}=2\qquad (n=2).
\end{equation}

\subsection{Type-$U3$}\label{typu3}

The $\mathbbm{Z}_4$ automorphism in  this case is
\begin{equation}\label{}
\Omega \circ\begin{pmatrix}
   A  & \Theta   \\
   \Psi   &  B \\
 \end{pmatrix}
 =
 \begin{pmatrix}
   -B^t  & i\Theta ^t  \\
    i\Psi ^t & -A^t  \\
 \end{pmatrix},
\end{equation}
which gives the following $\mathbbm{Z}_4$ decomposition:
\begin{eqnarray}\label{}
 \mathfrak{h}_2:&&
 B=A^t
 \nonumber \\
 \mathfrak{h}_1:&&
 \Theta^t =\Theta,\qquad \Psi^t =\Psi
 \nonumber \\
 \mathfrak{h}_3:&&
 \Theta ^t=-\Theta ,\qquad \Psi ^t=-\Psi .
\end{eqnarray}
This case is rather special, because $\mathfrak{h}_1$ and
$\mathfrak{h}_3$ have different dimensions:
$\dim\mathfrak{h}_1-\dim\mathfrak{h}_3=2n$, potentially leading to
the mismatch of the central charges of left and right movers. We
will see that this mismatch is precisely compensated by the
kappa-symmetry.

The equation~(\ref{comma}), that determines the rank of the
kappa-symmetry, becomes:
\begin{equation}\label{matrya}
 A\Theta =\Theta A^t,\qquad A^t\Psi =\Psi A.
\end{equation}
The solutions of this equation in symmetric matrices give
$\hat{N}_{\tilde{\kappa} }$, the number of solutions in
anti-symmetric matrices determines $\hat{N}_\kappa $. Without loss
of generality we can assume that $A$ is diagonal:
$A=\mathop{\mathrm{diag}}(a_1,\ldots ,a_n)$. Then,
\begin{equation}\label{}
 (a_i-a_j)\Theta _{ij}=0,\qquad (a_i-a_j)\Psi _{ij}=0.
\end{equation}
In general, all the eigenvalues are different, and the solutions are
diagonal matrix elements $\Theta_{ii} $ and $\Psi _{ii}$. All of
them belong to $\mathfrak{h}_1$, thus giving:
\begin{equation}\label{}
 \hat{N}_{\tilde{\kappa }}=2n,\qquad \hat{N}_\kappa =0.
\end{equation}
The kappa-symmetry eliminates the extra right-moving degrees of
freedom and reinstalls the balance of central charges: $c_L=c_R$.
The Virasoro condition does not impose any new constraints, since
the equation $\,{\rm tr}\,A^2=\,{\rm tr}\,B^2$ automatically holds
for any element of $\mathfrak{h}_2$.

\subsection{Type-$U4$}

The type-$U4$ cosets are defined for $PSU(n|n)$ with even $n$. We
can also assume that $n>2$, as $\mathfrak{h}_2$ is empty
for\footnote{Once can however consider the coset of $U(2|2)$ (or
even $U(1|1)$) instead of $PSU(2|2)$, then two bosonic directions
survive the $\mathbbm{Z}_4$ projection \cite{Stefanski:2007dp}.}
$n=2$. The $\mathbbm{Z}_4$ symmetry acts as
\begin{equation}\label{}
\Omega \circ\begin{pmatrix}
   A  & \Theta   \\
   \Psi   &  B \\
 \end{pmatrix}
 =
 \begin{pmatrix}
   JA^tJ  & -J\Psi ^tJ  \\
    J\Theta ^tJ & JB^tJ  \\
    \end{pmatrix},
\end{equation}
and leads to the $\mathbbm{Z}_4$ decomposition
\begin{eqnarray}\label{tipu4h2}
 \mathfrak{h}_2:&&
 A^t=-JAJ,\qquad B^t=-JBJ
 \\
 \mathfrak{h}_1:&&
 \Psi =-iJ\Theta ^tJ
 \nonumber \\
 \mathfrak{h}_3:&&
 \Psi =iJ\Theta ^tJ.
\end{eqnarray}

The second equation in (\ref{comma}) is a consequence of the first
for this coset. To find the rank of the kappa-symmetry, we just need
to solve
\begin{equation}\label{A=B}
 A\Theta =\Theta B.
\end{equation}
Non-trivial  solutions to this equation correspond to pairs of equal
eigenvalues of $A$ and $B$.

A matrix that satisfies the condition (\ref{tipu4h2}) can be
represented as a product of two anti-symmetric matrices: $A=-(AJ)J$,
and consequently has doubly-degenerate spectrum, by Stenzel theorem.
In addition, $A$ and $B$ are traceless and so have $n/2-1$
independent eigenvalues. In general these eigenvalues have no
reasons to coincide. Consequently, the coset has no extrinsic
kappa-symmetries:
\begin{equation}\label{}
 \hat{N}_\kappa =\hat{N}_{\tilde{\kappa }}=0.
\end{equation}

The Virasoro condition (\ref{virnull}) imposes the relationship on
the sums of squares of the eigenvalues of $A$ and $B$. In general
this is not enough to force them to coincide. The only exception is
the case of $n=4$, when $A$ and $B$ have just one independent
eigenvalue each: $\{\alpha ,\alpha -\alpha ,-\alpha \}$ and $\{\beta
,\beta ,-\beta ,-\beta \}$. The Virasoro condition means that
$\alpha =\pm\beta $. Then (\ref{A=B}) has an eight-dimensional space
of solutions, and
\begin{equation}\label{}
 N_\kappa =N_{\tilde{\kappa }}=8\qquad (n=4).
\end{equation}

\subsection{Type-$O1$}\label{typeo1}

Since $\Psi $ and $\Theta $  in $\mathfrak{osp}(2n+2|2n)$ are
related, for type-$O$ cosets we only need to solve one equation in
(\ref{comma}):
\begin{equation}\label{whattosolve}
 A\Theta =\Theta B,
\end{equation}
the other will automatically follow.

The $\mathbbm{Z}_4$ automorphism of the type-$O1$ cosets in the
supermatrix representation (\ref{osp2n2nmatrix}) acts as follows:
\begin{equation}\label{}
\Omega \circ\begin{pmatrix}
   A  & \Theta   \\
   \Psi   &  B \\
 \end{pmatrix}
 =
 \begin{pmatrix}
   I_pAI_p  & I_p\Theta J  \\
    -J\Psi I_p & -JBJ  \\
    \end{pmatrix}=\begin{pmatrix}
   I_pAI_p  & I_p\Theta J  \\
    -J\Psi I_p & -B^t  \\
    \end{pmatrix}.
\end{equation}
The associated $\mathbbm{Z}_4$ decomposition is
\begin{eqnarray}\label{}
 \mathfrak{h}_2:&&
 A=
 \begin{pmatrix}
   0_{p\times p}  & [A_1]_{p\times (2n+2-p)}  \\
   -[A^t_1]_{(2n+2-p)\times p}  & 0_{(2n+2-p)\times (2n+2-p)}  \\
 \end{pmatrix},
 \nonumber \\
 &&B=
 \begin{pmatrix}
   [B_1]_{n\times n}  & [B_2]_{n\times n}  \\
    [B_2]_{n\times n}  & -  [B_1]_{n\times n} \\
 \end{pmatrix}, B_i^t=B_i
 \nonumber \\
 \mathfrak{h}_1:&&
 \Theta =
 \begin{pmatrix}
   [\Theta_1]_{p\times n}  & -i[\Theta_1]_{p\times n}  \\
   [\Theta_2]_{(2n+2-p)\times n}  & i[\Theta_2]_{(2n+2-p)\times n}  \\
 \end{pmatrix}
 \nonumber \\
 \mathfrak{h}_3:&&
 \Theta =
 \begin{pmatrix}
   [\Theta_1]_{p\times n}  & i[\Theta_1]_{p\times n}  \\
   [\Theta_2]_{(2n+2-p)\times n}  & -i[\Theta_2]_{(2n+2-p)\times n}  \\
 \end{pmatrix}
\end{eqnarray}

The kappa-symmetry condition (\ref{whattosolve}) in
$\mathfrak{h}_{1/3}$ reduces to
\begin{equation}\label{}
 A_1\Theta _2=\Theta _1B_\mp,\qquad -A_1^t\Theta _1=\Theta _2B_\pm,
\end{equation}
where
\begin{equation}\label{}
 B_\pm=B_1\pm iB_2.
\end{equation}
Then $\Theta _2=-A_1^t\Theta _1B_\pm^{-1}$, and we are left with the
equation
\begin{equation}\label{}
 -A_1A_1^t\Theta _1=\Theta _1B_\mp B_\pm
\end{equation}
for the $p\times n$ matrix $\Theta _1$. In general this equation has
no solutions and, consequently, there will be no kappa-symmetries:
\begin{equation}\label{}
 \hat{N}_\kappa=\hat{N}_{\tilde{\kappa }} =0,
\end{equation}
because $-A_1A_1^t$ and $B_\mp B_\pm$ have different eigenvalues for
generic matrices $A_1$, $B_1$, $B_2$. The null condition
(\ref{virnull}) relates the sums of the eigenvalues, because
\begin{equation}\label{}
 \,{\rm tr}\,A^2=-2\,{\rm tr}\,A_1A_1^t,\qquad \,{\rm tr}\,B^2=2\,{\rm
 tr}\,B_\mp B_\pm.
\end{equation}
This is still insufficient for the eigenvalues to coincide, except
for the special case of $p=1$, $n=1$. Then both $-A_1A_1^t$ and
$B_\mp B_\pm$ are numbers rather than matrices, which must coincide
once the trace condition (\ref{virnull}) is imposed. We thus find
one solution in $\mathfrak{h}_1$ and one in $\mathfrak{h}_3$:
\begin{equation}\label{}
 N_\kappa =N_{\tilde{\kappa }}=1\qquad \left(p=1, n=1\right).
\end{equation}

There is also an extremely degenerate case of $p=0$, $n=1$. The
target space then is $AdS_2$, without any extra factors. There are
no propagating bosonic degrees of freedom. The number of
kappa-symmetries, and consequently the number of fermionic degrees
of freedom, depends on whether the string is left- or right-moving
in the target space. In one case, there is no kappa-symmetries, and
in the other case the kappa-symmetry removes all the fermions:
$N_\kappa =4=N_{\tilde{\kappa }}$. The string then is purely
topological.

\subsection{Type-$O2$}

The $\mathbbm{Z}_4$ automorphism in this case acts as
\begin{equation}\label{}
\Omega \circ\begin{pmatrix}
   A  & \Theta   \\
   \Psi   &  B \\
 \end{pmatrix}
 =
 \begin{pmatrix}
   -JAJ  & -J\Theta\,\mathbbm{1}\otimes I_p \\
    \mathbbm{1}\otimes I_p \Psi J & \mathbbm{1}\otimes I_pB\mathbbm{1}\otimes I_p  \\
    \end{pmatrix}.
\end{equation}
It is convenient to work in the basis in which
\begin{equation}\label{}
\mathbbm{1}\otimes I_p =
\begin{pmatrix}
  \mathbbm{1}_{2p\times 2p}  & 0  \\
   0 & \mathbbm{1}_{(2n-2p)\times (2n-2p)}  \\
\end{pmatrix}.
\end{equation}
The $\mathbbm{Z}_4$ decomposition in this basis takes the form (we
assume that $n-p\geq p$):
\begin{eqnarray}\label{}
 \mathfrak{h}_2:&&
 A=
 \begin{pmatrix}
   [A_1]_{(n+1)\times (n+1)}  & [A_2]_{(n+1)\times (n+1)}  \\
    [A_2]_{(n+1)\times (n+1)} &  -[A_1]_{(n+1)\times (n+1)} \\
 \end{pmatrix}, A_i^t=-A_i
 \nonumber \\
 &&B=
 \begin{pmatrix}
   0_{2p\times 2p}  & [B_1]_{2p\times (2n-2p)}  \\
   J_{(2n-2p)\times (2n-2p)}[B^t_1]_{(2n-2p)\times 2p}J_{2p\times 2p}  & 0_{(2n-2p)\times (2n-2p)}  \\
 \end{pmatrix}
 \nonumber \\
 \mathfrak{h}_1:&&
 \Theta =
 \begin{pmatrix}
   [\Theta _1]_{(n+1)\times 2p}  & [\Theta _2]_{(n+1)\times (2n-2p)}  \\
   -i[\Theta _1]_{(n+1)\times 2p}  &  i[\Theta _2]_{(n+1)\times (2n-2p)} \\
 \end{pmatrix}
 \nonumber \\
 \mathfrak{h}_3:&&
\Theta =
 \begin{pmatrix}
   [\Theta _1]_{(n+1)\times 2p}  & [\Theta _2]_{(n+1)\times (2n-2p)}  \\
   i[\Theta _1]_{(n+1)\times 2p}  &  -i[\Theta _2]_{(n+1)\times (2n-2p)} \\
 \end{pmatrix}
 \nonumber
\end{eqnarray}

The zero-mode equation (\ref{whattosolve}) in the
$\mathfrak{h}_{1/3}$ subspace boils down to  a system of two
equations for matrices $\Theta _1$, $\Theta _2$:
\begin{eqnarray}\label{Ath1}
 A_\mp\Theta _1&=&\Theta _2JB^t_1J
 \nonumber \\
 A_\pm\Theta _2&=&\Theta _1B_1,
\end{eqnarray}
where
\begin{equation}\label{}
 A_\pm=A_1\pm iA_2.
\end{equation}

We need to distinguish even and odd $n$. Consider first odd $n$. The
anti-symmetric matrices $A_\pm$ are then non-degenerate and we can
express $\Theta _2$ through $\Theta _1$: $\Theta _2=A_\pm^{-1}\Theta
_1B_1$, substitute the result into the first equation and get:
\begin{equation}\label{Anonz}
 A_\pm A_\mp\Theta _1=\Theta _1B_1JB_1^tJ.
\end{equation}
Both $A_\pm A_\mp$ and $(B_1JB_1^t)J$ are products of two
anti-symmetric matrices, their spectra are thus degenerate and
contain, respectively, $(n+1)/2$ and $p$ different eigenvalues:
$\alpha _1,\alpha _1,\ldots ,\alpha _{(n+1)/2},\alpha _{(n+1)/2}$
and $\beta _1,\beta _1,\ldots \beta _p,\beta _p$. These eigenvalues
are in general different. Hence there are no non-trivial solutions
for $\Theta _1$, and there are no kappa-symmetries:
\begin{equation}\label{}
 \hat{N}_{\kappa }=\hat{N}_{\tilde{\kappa }}=0 \qquad (n~{\rm odd}).
\end{equation}
Imposing the Virasoro constraints does not change the situation,
because the condition (\ref{virnull}) imposes just one constraint on
$\alpha _i$, $\beta _j$:
\begin{equation}\label{summyab}
\sum_{i}^{}\alpha _i=\sum_{j}^{}\beta _j.
\end{equation}
The only exception is the degenerate case of $n=1$, $p=0$, which is
completely analogous to the type-$O1$ coset with $n=1$, $p=0$,
discussed at the end of sec.~\ref{typeo1}.

If $n$ is even, the matrix $A_\pm$ has one zero eigenvalue:
\begin{equation}\label{}
 A_\pm v=0.
\end{equation}
This gives $2n-4p$ solutions of (\ref{Ath1}), in combination with
$2n-4p$ null eigenvalues of $JB^t_1J$:
\begin{equation}\label{}
 u_iJB_1^tJ=0,\qquad i=1,\ldots 2n-4p.
\end{equation}
The solutions due to the null eigenvalues are:
\begin{equation}\label{}
 \Theta _1=0,\qquad \Theta _2=v\otimes u_i.
\end{equation}
Potentially, there may also be solutions due to coincident non-zero
eigenvalues of $A$ and $B$, which are given by the equation
(\ref{Anonz}). This requires coincidence of some eigenvalues of
$A_\pm A_\pm$, $\alpha _1,\alpha _1,\ldots ,\alpha _{n/2},\alpha
_{n/2},0$, and $B_1JB_1^tJ$, $\beta _1,\beta _1,\ldots ,\beta
_p,\beta _p$. In general, this does not happen, and thus
\begin{equation}\label{}
 \hat{N}_\kappa =2n-4p=\hat{N}_{\tilde{\kappa }}\qquad (n~{\rm even}).
\end{equation}
But if we impose the Virasoro condition, the eigenvalues satisfy
(\ref{summyab}), and for $n=2$, $p=1$, the matrices $A_\pm A_\pm$
and $B_1JB_1^tJ$ have two pairs of coinciding eigenvalues, leading
to $4$ extra solutions:
\begin{equation}\label{}
 N_\kappa =4=N_{\tilde{\kappa }}\qquad (n=2,p=1).
\end{equation}

\subsection{Type-$Tu$}

The $\mathbbm{Z}_4$ generator of the the tensor-product models
(\ref{omegad}) acts on $\mathfrak{g}=\mathfrak{p}\oplus\mathfrak{p}$
as
\begin{equation}\label{}
 \Omega (X,Y)=(Y,(-1)^F\circ X),
\end{equation}
and gives the following $\mathbbm{Z}_4$ decomposition:
\begin{eqnarray}\label{}
 \mathfrak{h}_2:&&
 (X,-X), \qquad X\in\mathfrak{p}_{\rm bos}
 \nonumber \\
 \mathfrak{h}_1:&&
 (\Xi  ,-i\Xi  ), \qquad \Xi  \in\mathfrak{p}_{\rm ferm}
 \nonumber \\
 \mathfrak{h}_3:&&
 (\Xi  ,i\Xi  ),\qquad  \Xi \in\mathfrak{p}_{\rm ferm}.
\end{eqnarray}

The kernel of $\mathop{\mathrm{ad}}K$, $K=(X,-X)\in\mathfrak{h}_2$
in $\mathfrak{h}_{1/3}$ is determined by the equation
\begin{equation}\label{}
 ([X,\Xi ],\pm i[X,\Xi ])=(0,0)~\Longleftrightarrow~ [X,\Xi ]=0,
\end{equation}
where $\Xi $ is an odd (fermionic) element of
$\mathfrak{p}=\mathfrak{psu}(n|n)$. This reduces to (\ref{comma})
for generic $n\times n$ matrices $\Theta $ and $\Psi $. In  general,
$A$ and $B$ have different eigenvalues, and there will be no
solutions yielding
\begin{equation}\label{}
 \hat{N}_\kappa =0=\hat{N}_{\tilde{\kappa }}.
\end{equation}

The only exceptional case in which the null condition
(\ref{virnull}) makes a difference is $n=2$. Then there are two
solutions for each of the matrices $\Theta $ and $\Psi $, and thus
\begin{equation}\label{}
 N_\kappa =N_{\tilde{\kappa }}=4\qquad (n=2).
\end{equation}

\subsection{Type-$To$}

The matrices $A$ and $B$ in (\ref{comma}) never have common
eigenvalues for $\mathfrak{p}=\mathfrak{osp}(2n+2|2n)$, even if the
null condition (\ref{virnull}) is imposed. So,
\begin{equation}\label{}
 \hat{N}_\kappa =\hat{N}_{\tilde{\kappa }}=0,
\end{equation}
and there are no exceptional cases.

\bibliographystyle{nb}
\bibliography{refs}

\begin{thebibliography}{10}
\ifx\href\asklfhas\newcommand{\href}[2]{#2}\fi
\raggedright
\small
\parskip 0pt

\bibitem{Serganova}
V.~V.~Serganova,
\textit{``{Classification of real simple Lie superalgebras and symmetric
  superspaces}''},
\textsf{Funct.~Anal.~Appl.~17,~200~(1983)}.
%
%%CITATION = HEP-TH/9907200;%%
\bibitem{Berkovits:1999zq}
N.~Berkovits, M.~Bershadsky, T.~Hauer, S.~Zhukov and B.~Zwiebach,
\textit{``{Superstring theory on $AdS_2\times S^2$ as a coset
  supermanifold}''},
\textsf{Nucl.~Phys.~B567,~61~(2000)},
\href{http://arXiv.org/abs/hep-th/9907200}{\texttt{hep-th/9907200}}.
%
%%CITATION = HEP-TH 9805028;%%
\bibitem{Metsaev:1998it}
R.~R.~Metsaev and A.~A.~Tseytlin,
\textit{``Type IIB superstring action in $AdS_5 \times S^5$ background''},
\textsf{Nucl.~Phys.~B533,~109~(1998)},
\href{http://arXiv.org/abs/hep-th/9805028}{\texttt{hep-th/9805028}}.
%
%%CITATION = HEP-TH/0010104;%%
\bibitem{Roiban:2000yy}
R.~Roiban and W.~Siegel,
\textit{``{Superstrings on $AdS_5\times S^5$ supertwistor space}''},
\textsf{JHEP~0011,~024~(2000)},
\href{http://arXiv.org/abs/hep-th/0010104}{\texttt{hep-th/0010104}}.
%
%%CITATION = HEP-TH 0305116;%%
\bibitem{Bena:2003wd}
I.~Bena, J.~Polchinski and R.~Roiban,
\textit{``Hidden symmetries of the {$AdS_5\times S^5$} superstring''},
\textsf{Phys.~Rev.~D69,~046002~(2004)},
\href{http://arXiv.org/abs/hep-th/0305116}{\texttt{hep-th/0305116}}.
%
%%CITATION = NUPHA,B155,381;%%
\bibitem{Eichenherr:1979ci}
H.~Eichenherr and M.~Forger,
\textit{``{On the Dual Symmetry of the Nonlinear Sigma Models}''},
\textsf{Nucl.~Phys.~B155,~381~(1979)}.
%
%%CITATION = PHLTA,B59,79;%%
\bibitem{Polyakov:1975rr}
A.~M.~Polyakov,
\textit{``{Interaction of Goldstone Particles in Two-Dimensions. Applications
  to Ferromagnets and Massive Yang-Mills Fields}''},
\textsf{Phys.~Lett.~B59,~79~(1975)}.
%
%%CITATION = HEP-TH/0405106;%%
\bibitem{Polyakov:2004br}
A.~M.~Polyakov,
\textit{``{Conformal fixed points of unidentified gauge theories}''},
\textsf{Mod.~Phys.~Lett.~A19,~1649~(2004)},
\href{http://arXiv.org/abs/hep-th/0405106}{\texttt{hep-th/0405106}}.
%
%%CITATION = HEP-TH/0702083;%%
\bibitem{Adam:2007ws}
I.~Adam, A.~Dekel, L.~Mazzucato and Y.~Oz,
\textit{``{Integrability of type II superstrings on Ramond-Ramond backgrounds
  in various dimensions}''},
\textsf{JHEP~0706,~085~(2007)},
\href{http://arXiv.org/abs/hep-th/0702083}{\texttt{hep-th/0702083}}.
%
%%CITATION = PHLTA,B136,367;%%
\bibitem{Green:1983wt}
M.~B.~Green and J.~H.~Schwarz,
\textit{``{Covariant Description of Superstrings}''},
\textsf{Phys.~Lett.~B136,~367~(1984)}.
%
%%CITATION = HEP-TH/9809164;%%
\bibitem{Rahmfeld:1998zn}
J.~Rahmfeld and A.~Rajaraman,
\textit{``{The GS string action on $AdS_3\times S^3$ with Ramond-Ramond
  charge}''},
\textsf{Phys.~Rev.~D60,~064014~(1999)},
\href{http://arXiv.org/abs/hep-th/9809164}{\texttt{hep-th/9809164}}.
%
%%CITATION = HEP-TH/9812062;%%
\bibitem{Park:1998un}
J.~Park and S.-J.~Rey,
\textit{``{Green-Schwarz superstring on $AdS_3\times S^3$}''},
\textsf{JHEP~9901,~001~(1999)},
\href{http://arXiv.org/abs/hep-th/9812062}{\texttt{hep-th/9812062}}.
%
%%CITATION = HEP-TH/9906013;%%
\bibitem{Zhou:1999sm}
J.-G.~Zhou,
\textit{``{Super 0-brane and GS superstring actions on $AdS_2\times S^2$}''},
\textsf{Nucl.~Phys.~B559,~92~(1999)},
\href{http://arXiv.org/abs/hep-th/9906013}{\texttt{hep-th/9906013}}.
%
%%CITATION = HEP-TH/0011191;%%
\bibitem{Metsaev:2000mv}
R.~R.~Metsaev and A.~A.~Tseytlin,
\textit{``{Superparticle and superstring in $AdS_3\times S^3$ Ramond-Ramond
  background in light-cone gauge}''},
\textsf{J.~Math.~Phys.~42,~2987~(2001)},
\href{http://arXiv.org/abs/hep-th/0011191}{\texttt{hep-th/0011191}}.
%
%%CITATION = HEP-TH/0403024;%%
\bibitem{Verlinde:2004gt}
H.~L.~Verlinde,
\textit{``{Superstrings on AdS(2) and superconformal matrix quantum
  mechanics}''},
\href{http://arXiv.org/abs/hep-th/0403024}{\texttt{hep-th/0403024}}.
%
%%CITATION = HEP-TH/0503089;%%
\bibitem{Chen:2005uj}
B.~Chen, Y.-L.~He, P.~Zhang and X.-C.~Song,
\textit{``{Flat currents of the Green-Schwarz superstrings in $AdS_5\times S^1$
  and $AdS_3\times S^3$ backgrounds}''},
\textsf{Phys.~Rev.~D71,~086007~(2005)},
\href{http://arXiv.org/abs/hep-th/0503089}{\texttt{hep-th/0503089}}.
%
%%CITATION = 0804.1831;%%
\bibitem{Hatsuda:2008xa}
M.~Hatsuda and Y.~Michishita,
\textit{``{Kappa symmetric $OSp(2|2)$ WZNW model}''},
\textsf{JHEP~,~049~(2008)},
\href{http://arXiv.org/abs/0804.1831}{\texttt{0804.1831}}.
%
%%CITATION = 0806.4940;%%
\bibitem{Arutyunov:2008if}
G.~Arutyunov and S.~Frolov,
\textit{``{Superstrings on $AdS_4 \times CP^3$ as a Coset Sigma-model}''},
\textsf{JHEP~0809,~129~(2008)},
\href{http://arXiv.org/abs/0806.4940}{\texttt{0806.4940}}.
%
%%CITATION = 0806.4948;%%
\bibitem{Stefanski:2008ik}
j.~Stefanski,~B.,
\textit{``{Green-Schwarz action for Type IIA strings on $AdS_4\times CP^3$}''},
\textsf{Nucl.~Phys.~B808,~80~(2009)},
\href{http://arXiv.org/abs/0806.4948}{\texttt{0806.4948}}.
%
%%CITATION = 0912.1723;%%
\bibitem{Babichenko:2009dk}
A.~Babichenko, B.~Stefanski and K.~Zarembo,
\textit{``{Integrability and the AdS(3)/CFT(2) correspondence}''},
\href{http://arXiv.org/abs/0912.1723}{\texttt{0912.1723}}.
%
%%CITATION = HEP-TH/0210064;%%
\bibitem{Vallilo:2002mh}
B.~C.~Vallilo,
\textit{``{One loop conformal invariance of the superstring in an $AdS_5\times
  S^5$ background}''},
\textsf{JHEP~0212,~042~(2002)},
\href{http://arXiv.org/abs/hep-th/0210064}{\texttt{hep-th/0210064}}.
%
%%CITATION = HEP-TH/0512250;%%
\bibitem{Kagan:2005wt}
D.~Kagan and C.~A.~S.~Young,
\textit{``{Conformal Sigma-Models on Supercoset Targets}''},
\textsf{Nucl.~Phys.~B745,~109~(2006)},
\href{http://arXiv.org/abs/hep-th/0512250}{\texttt{hep-th/0512250}}.
%
%%CITATION = HEP-TH/0607076;%%
\bibitem{Puletti:2006vb}
V.~G.~M.~Puletti,
\textit{``{Operator product expansion for pure spinor superstring on
  $AdS_5\times S^5$}''},
\textsf{JHEP~0610,~057~(2006)},
\href{http://arXiv.org/abs/hep-th/0607076}{\texttt{hep-th/0607076}}.
%
%%CITATION = 0906.4572;%%
\bibitem{Mazzucato:2009fv}
L.~Mazzucato and B.~C.~Vallilo,
\textit{``{On the Non-renormalization of the AdS Radius}''},
\textsf{JHEP~0909,~056~(2009)},
\href{http://arXiv.org/abs/0906.4572}{\texttt{0906.4572}}.
%
%%CITATION = NUPHA,B284,365;%%
\bibitem{Carlip:1986cz}
S.~Carlip,
\textit{``{Heterotic string path integrals with the Green-Schwarz covariant
  action}''},
\textsf{Nucl.~Phys.~B284,~365~(1987)}.
%
%%CITATION = IMPAE,A3,1943;%%
\bibitem{Kallosh:1988wv}
R.~Kallosh and A.~Y.~Morozov,
\textit{``{Green-Schwarz Action and Loop Calculations for Superstring}''},
\textsf{Int.~J.~Mod.~Phys.~A3,~1943~(1988)}.
%
%%CITATION = NUPHA,B323,330;%%
\bibitem{Wiegmann:1989md}
P.~B.~Wiegmann,
\textit{``{Extrinsic Geometry Of Superstrings}''},
\textsf{Nucl.~Phys.~B323,~330~(1989)}.
%
%%CITATION = HEP-TH/9902098;%%
\bibitem{Berkovits:1999im}
N.~Berkovits, C.~Vafa and E.~Witten,
\textit{``{Conformal field theory of AdS background with Ramond-Ramond
  flux}''},
\textsf{JHEP~9903,~018~(1999)},
\href{http://arXiv.org/abs/hep-th/9902098}{\texttt{hep-th/9902098}}.
%
%%CITATION = HEP-TH/9902180;%%
\bibitem{Bershadsky:1999hk}
M.~Bershadsky, S.~Zhukov and A.~Vaintrob,
\textit{``{$PSL(n|n)$ sigma model as a conformal field theory}''},
\textsf{Nucl.~Phys.~B559,~205~(1999)},
\href{http://arXiv.org/abs/hep-th/9902180}{\texttt{hep-th/9902180}}.
%
%%CITATION = HEP-TH/0106124;%%
\bibitem{Read:2001pz}
N.~Read and H.~Saleur,
\textit{``{Exact spectra of conformal supersymmetric nonlinear sigma models in
  two dimensions}''},
\textsf{Nucl.~Phys.~B613,~409~(2001)},
\href{http://arXiv.org/abs/hep-th/0106124}{\texttt{hep-th/0106124}}.
%
%%CITATION = HEP-TH/0611214;%%
\bibitem{Babichenko:2006uc}
A.~Babichenko,
\textit{``{Conformal invariance and quantum integrability of sigma models on
  symmetric superspaces}''},
\textsf{Phys.~Lett.~B648,~254~(2007)},
\href{http://arXiv.org/abs/hep-th/0611214}{\texttt{hep-th/0611214}}.
%
%%CITATION = CMPHA,53,31;%%
\bibitem{Kac:1977qb}
V.~G.~Kac,
\textit{``{A Sketch of Lie Superalgebra Theory}''},
\textsf{Commun.~Math.~Phys.~53,~31~(1977)}.
%
%%CITATION = HEP-TH/9607161;%%
\bibitem{Frappat:1996pb}
L.~Frappat, P.~Sorba and A.~Sciarrino,
\textit{``{Dictionary on Lie superalgebras}''},
\href{http://arXiv.org/abs/hep-th/9607161}{\texttt{hep-th/9607161}}.
%
%%CITATION = PHLTA,B236,255;%%
\bibitem{Kraemmer:1989af}
U.~Kraemmer and A.~Rebhan,
\textit{``{Anomalous anomalies in the Carlip-Kallosh quantization of the
  Green-Schwarz superstring}''},
\textsf{Phys.~Lett.~B236,~255~(1990)}.
%
%%CITATION = PHLTA,B253,67;%%
\bibitem{Bastianelli:1990xn}
F.~Bastianelli, P.~van~Nieuwenhuizen and A.~Van~Proeyen,
\textit{``{Superstring anomalies in the semilight cone gauge}''},
\textsf{Phys.~Lett.~B253,~67~(1991)}.
%
%%CITATION = PHLTA,B273,47;%%
\bibitem{Porrati:1991ts}
M.~Porrati and P.~van~Nieuwenhuizen,
\textit{``{Absence of world sheet and space-time anomalies in the
  semicovariantly quantized heterotic string}''},
\textsf{Phys.~Lett.~B273,~47~(1991)}.
%
%%CITATION = NUPHA,B363,573;%%
\bibitem{Bellucci:1991hy}
S.~Bellucci and R.~N.~Oerter,
\textit{``{Weyl invariance of the Green-Schwarz heterotic sigma model}''},
\textsf{Nucl.~Phys.~B363,~573~(1991)}.
%
%%CITATION = 0903.4277;%%
\bibitem{Ashok:2009xx}
S.~K.~Ashok, R.~Benichou and J.~Troost,
\textit{``{Conformal Current Algebra in Two Dimensions}''},
\textsf{JHEP~0906,~017~(2009)},
\href{http://arXiv.org/abs/0903.4277}{\texttt{0903.4277}}.
%
%%CITATION = 0704.1460;%%
\bibitem{Stefanski:2007dp}
B.~Stefanski,~Jr.,
\textit{``{Landau-Lifshitz sigma-models, fermions and the AdS/CFT
  correspondence}''},
\textsf{JHEP~0707,~009~(2007)},
\href{http://arXiv.org/abs/0704.1460}{\texttt{0704.1460}}.
%
%%CITATION = 0811.1566;%%
\bibitem{Gomis:2008jt}
J.~Gomis, D.~Sorokin and L.~Wulff,
\textit{``{The complete $AdS_4\times CP^3$ superspace for the type IIA
  superstring and D-branes}''},
\href{http://arXiv.org/abs/0811.1566}{\texttt{0811.1566}}.
%
%%CITATION = HEP-TH/9809145;%%
\bibitem{Pesando:1998wm}
I.~Pesando,
\textit{``{The GS type IIB superstring action on $AdS_3\times S^3\times
  T^4$}''},
\textsf{JHEP~9902,~007~(1999)},
\href{http://arXiv.org/abs/hep-th/9809145}{\texttt{hep-th/9809145}}.
%
%%CITATION = 0908.0878;%%
\bibitem{Candu:2009ep}
C.~Candu, V.~Mitev, T.~Quella, H.~Saleur and V.~Schomerus,
\textit{``{The Sigma Model on Complex Projective Superspaces}''},
\textsf{JHEP~1002,~015~(2010)},
\href{http://arXiv.org/abs/0908.0878}{\texttt{0908.0878}}.
%
%%CITATION = 1001.1344;%%
\bibitem{Candu:2010yg}
C.~Candu, T.~Creutzig, V.~Mitev and V.~Schomerus,
\textit{``{Cohomological Reduction of Sigma Models}''},
\href{http://arXiv.org/abs/1001.1344}{\texttt{1001.1344}}.
%
%%CITATION = HEP-TH 0402207;%%
\bibitem{Kazakov:2004qf}
V.~A.~Kazakov, A.~Marshakov, J.~A.~Minahan and K.~Zarembo,
\textit{``Classical/quantum integrability in AdS/CFT''},
\textsf{JHEP~0405,~024~(2004)},
\href{http://arXiv.org/abs/hep-th/0402207}{\texttt{hep-th/0402207}}.
%
%%CITATION = HEP-TH/0502226;%%
\bibitem{Beisert:2005bm}
N.~Beisert, V.~A.~Kazakov, K.~Sakai and K.~Zarembo,
\textit{``{The algebraic curve of classical superstrings on $AdS_5\times
  S^5$}''},
\textsf{Commun.~Math.~Phys.~263,~659~(2006)},
\href{http://arXiv.org/abs/hep-th/0502226}{\texttt{hep-th/0502226}}.
%
%%CITATION = 0807.0437;%%
\bibitem{Gromov:2008bz}
N.~Gromov and P.~Vieira,
\textit{``{The AdS4/CFT3 algebraic curve}''},
\textsf{JHEP~0902,~040~(2009)},
\href{http://arXiv.org/abs/0807.0437}{\texttt{0807.0437}}.
%
%%CITATION = HEP-TH/0511082;%%
\bibitem{Beisert:2005tm}
N.~Beisert,
\textit{``{The $su(2|2)$ dynamic S-matrix}''},
\textsf{Adv.~Theor.~Math.~Phys.~12,~945~(2008)},
\href{http://arXiv.org/abs/hep-th/0511082}{\texttt{hep-th/0511082}}.
%
%%CITATION = 0807.1924;%%
\bibitem{Ahn:2008aa}
C.~Ahn and R.~I.~Nepomechie,
\textit{``{N=6 super Chern-Simons theory S-matrix and all-loop Bethe ansatz
  equations}''},
\textsf{JHEP~0809,~010~(2008)},
\href{http://arXiv.org/abs/0807.1924}{\texttt{0807.1924}}.
%
%%CITATION = HEP-TH 0504190;%%
\bibitem{Beisert:2005fw}
N.~Beisert and M.~Staudacher,
\textit{``Long-range $PSU(2,2|4)$ Bethe ansaetze for gauge theory and
  strings''},
\textsf{Nucl.~Phys.~B727,~1~(2005)},
\href{http://arXiv.org/abs/hep-th/0504190}{\texttt{hep-th/0504190}}.
%
%%CITATION = 0807.0777;%%
\bibitem{Gromov:2008qe}
N.~Gromov and P.~Vieira,
\textit{``{The all loop AdS4/CFT3 Bethe ansatz}''},
\textsf{JHEP~0901,~016~(2009)},
\href{http://arXiv.org/abs/0807.0777}{\texttt{0807.0777}}.
%
%%CITATION = 0901.3753;%%
\bibitem{Gromov:2009tv}
N.~Gromov, V.~Kazakov and P.~Vieira,
\textit{``{Integrability for the Full Spectrum of Planar AdS/CFT}''},
\href{http://arXiv.org/abs/0901.3753}{\texttt{0901.3753}}.
%
%%CITATION = 0902.3930;%%
\bibitem{Bombardelli:2009ns}
D.~Bombardelli, D.~Fioravanti and R.~Tateo,
\textit{``{Thermodynamic Bethe Ansatz for planar AdS/CFT: a proposal}''},
\textsf{J.~Phys.~A42,~375401~(2009)},
\href{http://arXiv.org/abs/0902.3930}{\texttt{0902.3930}}.
%
%%CITATION = 0902.4458;%%
\bibitem{Gromov:2009bc}
N.~Gromov, V.~Kazakov, A.~Kozak and P.~Vieira,
\textit{``{Integrability for the Full Spectrum of Planar AdS/CFT II}''},
\href{http://arXiv.org/abs/0902.4458}{\texttt{0902.4458}}.
%
%%CITATION = 0903.0141;%%
\bibitem{Arutyunov:2009ur}
G.~Arutyunov and S.~Frolov,
\textit{``{Thermodynamic Bethe Ansatz for the $AdS_5 \times S^5$ Mirror
  Model}''},
\textsf{JHEP~0905,~068~(2009)},
\href{http://arXiv.org/abs/0903.0141}{\texttt{0903.0141}}.
%
%%CITATION = 0912.4715;%%
\bibitem{Bombardelli:2009xz}
D.~Bombardelli, D.~Fioravanti and R.~Tateo,
\textit{``{TBA and Y-system for planar $AdS_4/CFT_3$}''},
\href{http://arXiv.org/abs/0912.4715}{\texttt{0912.4715}}.
%
%%CITATION = 0912.4911;%%
\bibitem{Gromov:2009at}
N.~Gromov and F.~Levkovich-Maslyuk,
\textit{``{Y-system, TBA and Quasi-Classical strings in $AdS_4\times CP^3$}''},
\href{http://arXiv.org/abs/0912.4911}{\texttt{0912.4911}}.
%
\bibitem{Stenzel}
H.~Stenzel,
\textit{``{\"{U}ber die Darstellbarkeit einer Matrix als Produkt von zwei
  symmetrischer Matrizen, als Produkt von zwei alternierenden Matrizen und als
  Produkt von einer symmetrischen und einer alternierenden Matrix}''},
\textsf{Math.~Z.~15,~1~(1922)}.
%
\bibitem{Ikramov}
K.~D.~Ikramov and H.~Fassbender,
\textit{``{On the product of two skew-Hamiltonian matrices or two
  skew-symmetric matrices}''},
\textsf{Zap.~Nauchn.~Sem.~POMI~359,~45~(2008)}.
%
\end{thebibliography}

\end{document}